\documentclass[letterpaper,compsoc,twoside]{IEEEtran}
\usepackage{fixltx2e} 
\usepackage{cmap} 
\usepackage{}
\usepackage{ifthen}
\usepackage[T1]{fontenc}
\usepackage[utf8]{inputenc}

\usepackage[font={small,it},labelfont=bf]{caption}
\usepackage{float}

\pdfoutput=1
\usepackage{scipy}
\makeatletter
\def\PY@reset{\let\PY@it=\relax \let\PY@bf=\relax%
    \let\PY@ul=\relax \let\PY@tc=\relax%
    \let\PY@bc=\relax \let\PY@ff=\relax}
\def\PY@tok#1{\csname PY@tok@#1\endcsname}
\def\PY@toks#1+{\ifx\relax#1\empty\else%
    \PY@tok{#1}\expandafter\PY@toks\fi}
\def\PY@do#1{\PY@bc{\PY@tc{\PY@ul{%
    \PY@it{\PY@bf{\PY@ff{#1}}}}}}}
\def\PY#1#2{\PY@reset\PY@toks#1+\relax+\PY@do{#2}}

\expandafter\def\csname PY@tok@gd\endcsname{\def\PY@tc##1{\textcolor[rgb]{0.63,0.00,0.00}{##1}}}
\expandafter\def\csname PY@tok@gu\endcsname{\let\PY@bf=\textbf\def\PY@tc##1{\textcolor[rgb]{0.50,0.00,0.50}{##1}}}
\expandafter\def\csname PY@tok@gt\endcsname{\def\PY@tc##1{\textcolor[rgb]{0.00,0.27,0.87}{##1}}}
\expandafter\def\csname PY@tok@gs\endcsname{\let\PY@bf=\textbf}
\expandafter\def\csname PY@tok@gr\endcsname{\def\PY@tc##1{\textcolor[rgb]{1.00,0.00,0.00}{##1}}}
\expandafter\def\csname PY@tok@cm\endcsname{\let\PY@it=\textit\def\PY@tc##1{\textcolor[rgb]{0.25,0.50,0.56}{##1}}}
\expandafter\def\csname PY@tok@vg\endcsname{\def\PY@tc##1{\textcolor[rgb]{0.73,0.38,0.84}{##1}}}
\expandafter\def\csname PY@tok@vi\endcsname{\def\PY@tc##1{\textcolor[rgb]{0.73,0.38,0.84}{##1}}}
\expandafter\def\csname PY@tok@mh\endcsname{\def\PY@tc##1{\textcolor[rgb]{0.13,0.50,0.31}{##1}}}
\expandafter\def\csname PY@tok@cs\endcsname{\def\PY@tc##1{\textcolor[rgb]{0.25,0.50,0.56}{##1}}\def\PY@bc##1{\setlength{\fboxsep}{0pt}\colorbox[rgb]{1.00,0.94,0.94}{\strut ##1}}}
\expandafter\def\csname PY@tok@ge\endcsname{\let\PY@it=\textit}
\expandafter\def\csname PY@tok@vc\endcsname{\def\PY@tc##1{\textcolor[rgb]{0.73,0.38,0.84}{##1}}}
\expandafter\def\csname PY@tok@il\endcsname{\def\PY@tc##1{\textcolor[rgb]{0.13,0.50,0.31}{##1}}}
\expandafter\def\csname PY@tok@go\endcsname{\def\PY@tc##1{\textcolor[rgb]{0.20,0.20,0.20}{##1}}}
\expandafter\def\csname PY@tok@cp\endcsname{\def\PY@tc##1{\textcolor[rgb]{0.00,0.44,0.13}{##1}}}
\expandafter\def\csname PY@tok@gi\endcsname{\def\PY@tc##1{\textcolor[rgb]{0.00,0.63,0.00}{##1}}}
\expandafter\def\csname PY@tok@gh\endcsname{\let\PY@bf=\textbf\def\PY@tc##1{\textcolor[rgb]{0.00,0.00,0.50}{##1}}}
\expandafter\def\csname PY@tok@ni\endcsname{\let\PY@bf=\textbf\def\PY@tc##1{\textcolor[rgb]{0.84,0.33,0.22}{##1}}}
\expandafter\def\csname PY@tok@nl\endcsname{\let\PY@bf=\textbf\def\PY@tc##1{\textcolor[rgb]{0.00,0.13,0.44}{##1}}}
\expandafter\def\csname PY@tok@nn\endcsname{\let\PY@bf=\textbf\def\PY@tc##1{\textcolor[rgb]{0.05,0.52,0.71}{##1}}}
\expandafter\def\csname PY@tok@no\endcsname{\def\PY@tc##1{\textcolor[rgb]{0.38,0.68,0.84}{##1}}}
\expandafter\def\csname PY@tok@na\endcsname{\def\PY@tc##1{\textcolor[rgb]{0.25,0.44,0.63}{##1}}}
\expandafter\def\csname PY@tok@nb\endcsname{\def\PY@tc##1{\textcolor[rgb]{0.00,0.44,0.13}{##1}}}
\expandafter\def\csname PY@tok@nc\endcsname{\let\PY@bf=\textbf\def\PY@tc##1{\textcolor[rgb]{0.05,0.52,0.71}{##1}}}
\expandafter\def\csname PY@tok@nd\endcsname{\let\PY@bf=\textbf\def\PY@tc##1{\textcolor[rgb]{0.33,0.33,0.33}{##1}}}
\expandafter\def\csname PY@tok@ne\endcsname{\def\PY@tc##1{\textcolor[rgb]{0.00,0.44,0.13}{##1}}}
\expandafter\def\csname PY@tok@nf\endcsname{\def\PY@tc##1{\textcolor[rgb]{0.02,0.16,0.49}{##1}}}
\expandafter\def\csname PY@tok@si\endcsname{\let\PY@it=\textit\def\PY@tc##1{\textcolor[rgb]{0.44,0.63,0.82}{##1}}}
\expandafter\def\csname PY@tok@s2\endcsname{\def\PY@tc##1{\textcolor[rgb]{0.25,0.44,0.63}{##1}}}
\expandafter\def\csname PY@tok@nt\endcsname{\let\PY@bf=\textbf\def\PY@tc##1{\textcolor[rgb]{0.02,0.16,0.45}{##1}}}
\expandafter\def\csname PY@tok@nv\endcsname{\def\PY@tc##1{\textcolor[rgb]{0.73,0.38,0.84}{##1}}}
\expandafter\def\csname PY@tok@s1\endcsname{\def\PY@tc##1{\textcolor[rgb]{0.25,0.44,0.63}{##1}}}
\expandafter\def\csname PY@tok@ch\endcsname{\let\PY@it=\textit\def\PY@tc##1{\textcolor[rgb]{0.25,0.50,0.56}{##1}}}
\expandafter\def\csname PY@tok@m\endcsname{\def\PY@tc##1{\textcolor[rgb]{0.13,0.50,0.31}{##1}}}
\expandafter\def\csname PY@tok@gp\endcsname{\let\PY@bf=\textbf\def\PY@tc##1{\textcolor[rgb]{0.78,0.36,0.04}{##1}}}
\expandafter\def\csname PY@tok@sh\endcsname{\def\PY@tc##1{\textcolor[rgb]{0.25,0.44,0.63}{##1}}}
\expandafter\def\csname PY@tok@ow\endcsname{\let\PY@bf=\textbf\def\PY@tc##1{\textcolor[rgb]{0.00,0.44,0.13}{##1}}}
\expandafter\def\csname PY@tok@sx\endcsname{\def\PY@tc##1{\textcolor[rgb]{0.78,0.36,0.04}{##1}}}
\expandafter\def\csname PY@tok@bp\endcsname{\def\PY@tc##1{\textcolor[rgb]{0.00,0.44,0.13}{##1}}}
\expandafter\def\csname PY@tok@c1\endcsname{\let\PY@it=\textit\def\PY@tc##1{\textcolor[rgb]{0.25,0.50,0.56}{##1}}}
\expandafter\def\csname PY@tok@o\endcsname{\def\PY@tc##1{\textcolor[rgb]{0.40,0.40,0.40}{##1}}}
\expandafter\def\csname PY@tok@kc\endcsname{\let\PY@bf=\textbf\def\PY@tc##1{\textcolor[rgb]{0.00,0.44,0.13}{##1}}}
\expandafter\def\csname PY@tok@c\endcsname{\let\PY@it=\textit\def\PY@tc##1{\textcolor[rgb]{0.25,0.50,0.56}{##1}}}
\expandafter\def\csname PY@tok@mf\endcsname{\def\PY@tc##1{\textcolor[rgb]{0.13,0.50,0.31}{##1}}}
\expandafter\def\csname PY@tok@err\endcsname{\def\PY@bc##1{\setlength{\fboxsep}{0pt}\fcolorbox[rgb]{1.00,0.00,0.00}{1,1,1}{\strut ##1}}}
\expandafter\def\csname PY@tok@mb\endcsname{\def\PY@tc##1{\textcolor[rgb]{0.13,0.50,0.31}{##1}}}
\expandafter\def\csname PY@tok@ss\endcsname{\def\PY@tc##1{\textcolor[rgb]{0.32,0.47,0.09}{##1}}}
\expandafter\def\csname PY@tok@sr\endcsname{\def\PY@tc##1{\textcolor[rgb]{0.14,0.33,0.53}{##1}}}
\expandafter\def\csname PY@tok@mo\endcsname{\def\PY@tc##1{\textcolor[rgb]{0.13,0.50,0.31}{##1}}}
\expandafter\def\csname PY@tok@kd\endcsname{\let\PY@bf=\textbf\def\PY@tc##1{\textcolor[rgb]{0.00,0.44,0.13}{##1}}}
\expandafter\def\csname PY@tok@mi\endcsname{\def\PY@tc##1{\textcolor[rgb]{0.13,0.50,0.31}{##1}}}
\expandafter\def\csname PY@tok@kn\endcsname{\let\PY@bf=\textbf\def\PY@tc##1{\textcolor[rgb]{0.00,0.44,0.13}{##1}}}
\expandafter\def\csname PY@tok@cpf\endcsname{\let\PY@it=\textit\def\PY@tc##1{\textcolor[rgb]{0.25,0.50,0.56}{##1}}}
\expandafter\def\csname PY@tok@kr\endcsname{\let\PY@bf=\textbf\def\PY@tc##1{\textcolor[rgb]{0.00,0.44,0.13}{##1}}}
\expandafter\def\csname PY@tok@s\endcsname{\def\PY@tc##1{\textcolor[rgb]{0.25,0.44,0.63}{##1}}}
\expandafter\def\csname PY@tok@kp\endcsname{\def\PY@tc##1{\textcolor[rgb]{0.00,0.44,0.13}{##1}}}
\expandafter\def\csname PY@tok@w\endcsname{\def\PY@tc##1{\textcolor[rgb]{0.73,0.73,0.73}{##1}}}
\expandafter\def\csname PY@tok@kt\endcsname{\def\PY@tc##1{\textcolor[rgb]{0.56,0.13,0.00}{##1}}}
\expandafter\def\csname PY@tok@sc\endcsname{\def\PY@tc##1{\textcolor[rgb]{0.25,0.44,0.63}{##1}}}
\expandafter\def\csname PY@tok@sb\endcsname{\def\PY@tc##1{\textcolor[rgb]{0.25,0.44,0.63}{##1}}}
\expandafter\def\csname PY@tok@k\endcsname{\let\PY@bf=\textbf\def\PY@tc##1{\textcolor[rgb]{0.00,0.44,0.13}{##1}}}
\expandafter\def\csname PY@tok@se\endcsname{\let\PY@bf=\textbf\def\PY@tc##1{\textcolor[rgb]{0.25,0.44,0.63}{##1}}}
\expandafter\def\csname PY@tok@sd\endcsname{\let\PY@it=\textit\def\PY@tc##1{\textcolor[rgb]{0.25,0.44,0.63}{##1}}}

\def\PYZus{\char`\_}
\def\PYZob{\char`\{}
\def\PYZcb{\char`\}}

\def\PYZsh{\char`\#}

\def\PYZhy{\char`\-}
\def\PYZsq{\char`\'}
\def\PYZdq{\char`\"}

\makeatother

\providecommand*{\DUprovidelength}[2]{
  \ifthenelse{\isundefined{#1}}{\newlength{#1}\setlength{#1}{#2}}{}
}
\providecommand*{\DUfootnotemark}[3]{%
  \raisebox{1em}{\hypertarget{#1}{}}%
  \hyperlink{#2}{\textsuperscript{#3}}%
}
\providecommand{\DUfootnotetext}[4]{%
  \begingroup%
  \renewcommand{\thefootnote}{%
    \protect\raisebox{1em}{\protect\hypertarget{#1}{}}%
    \protect\hyperlink{#2}{#3}}%
  \footnotetext{#4}%
  \endgroup%
}

\providecommand*{\DUrole}[2]{%
  \ifcsname DUrole#1\endcsname%
    \csname DUrole#1\endcsname{#2}%
  \else
    \ifcsname docutilsrole#1\endcsname%
      \csname docutilsrole#1\endcsname{#2}%
    \else%
      #2%
    \fi%
  \fi%
}

\DUprovidelength{\DUlineblockindent}{2.5em}
\ifthenelse{\isundefined{\DUlineblock}}{
  \newenvironment{DUlineblock}[1]{%
    \list{}{\setlength{\partopsep}{\parskip}
            \addtolength{\partopsep}{\baselineskip}
            \setlength{\topsep}{0pt}
            \setlength{\itemsep}{0.15\baselineskip}
            \setlength{\parsep}{0pt}
            \setlength{\leftmargin}{#1}}
    \raggedright
  }
  {\endlist}
}{}

\providecommand*{\DUroletitlereference}[1]{\textsl{#1}}

\ifthenelse{\isundefined{\hypersetup}}{
  \usepackage[colorlinks=true,linkcolor=blue,urlcolor=blue]{hyperref}
  \urlstyle{same} 
}{}

\begin{document}
\newcounter{footnotecounter}\title{Garbage Collection in JyNI – How to bridge Mark/Sweep and Reference Counting GC}\author{Stefan Richthofer$^{\setcounter{footnotecounter}{1}\fnsymbol{footnotecounter}\setcounter{footnotecounter}{2}\fnsymbol{footnotecounter}}$%
          \setcounter{footnotecounter}{1}\thanks{\fnsymbol{footnotecounter} %
          Corresponding author: \protect\href{mailto:stefan.richthofer@gmx.de}{stefan.richthofer@gmx.de}}\setcounter{footnotecounter}{2}\thanks{\fnsymbol{footnotecounter} Institute for Neural Computation, Ruhr-Universität Bochum}\thanks{%

          \noindent%
          Copyright\,\copyright\,2015 Stefan Richthofer. This is an open-access article distributed under the terms of the Creative Commons Attribution License, which permits unrestricted use, distribution, and reproduction in any medium, provided the original author and source are credited. http://creativecommons.org/licenses/by/3.0/%
        }}\maketitle
          \renewcommand{\leftmark}{PROC. OF THE 8th EUR. CONF. ON PYTHON IN SCIENCE (EUROSCIPY 2015)}
          \renewcommand{\rightmark}{GARBAGE COLLECTION IN JYNI – HOW TO BRIDGE MARK/SWEEP AND REFERENCE COUNTING GC}

\setcounter{page}{39}
\newcommand*{\docutilsroleref}{\ref}
\newcommand*{\docutilsrolelabel}{\label}
\AtEndDocument{\cleardoublepage}
\begin{abstract}Jython is a Java-based Python implementation and the most seamless way to
integrate Python and Java. It achieves high efficiency by compiling
Python code to Java bytecode and thus letting Java's JIT optimize it – an
approach that enables Python code to call Java functions or to subclass
Java classes. It enables Python code to leverage Java's
multithreading features and utilizes Java's built-in garbage collection (GC).
However, it currently does not support CPython's C-API and thus does not
support native extensions like NumPy and SciPy. Since most scientific code
depends on such extensions, it is not runnable with Jython.

Jython Native Interface (JyNI) is a compatibility layer that aims to provide
CPython's native C extension API on top of Jython. JyNI is implemented using
the Java Native Interface (JNI) and its native part is designed to be binary
compatible with existing extension builds. This means Jython can import the
original C extensions, i.e. the same .dll- or .so-files that CPython would use.

For various reasons, implementing CPython's C-API is not an easy task.
Just to name a few issues – it offers macros to access CPython internals,
uses a global interpreter lock (GIL) in contrast to Jython and lets extensions
perform reference-counting-based GC, which is incompatible
with Java's GC-approach. For each of the arising issues, JyNI proposes a
feasible solution; most remarkably it emulates CPython's reference-counting
GC on top of Java's mark-and-sweep-based approach (taking
care of adjacent concepts like finalizers and weak references and their
interference with Jython). (Note that there are vague considerations around
to switch to mark-and-sweep-based GC in a future CPython too; cf. \cite{PY3_PL15}. So this
algorithm might one day be even relevant to CPython in terms of running
legacy modules.)

This work's main purpose is to describe the algorithm JyNI uses to support
GC in detail, also covering weak references and testing native memory management.
Beside this we give a comprehension of JyNI's general design and briefly describe
how it deals with CPython's GIL. Finally we provide runnable code
examples, e.g. a demonstration of JyNI's support for the ctypes extension and a
first experimental import of NumPy.\end{abstract}\begin{IEEEkeywords}Jython, Java, Python, CPython, extensions, integration, JNI, native, NumPy, C-API, SciPy, GC\end{IEEEkeywords}

\section{Introduction%
  \label{introduction}%
}

As interpreter based languages, Python and Java both depend on native language bindings/extensions in many scenarios. Especially scientific code mostly relies on NumPy or native interfaces to some computation- or control-framework that connects Python to problem-specific hardware or libraries – a fact that usually ties this kind of code to CPython.
Developing and maintaining such bindings is usually a difficult and error-prone task. One major goal of the JyNI-project is to let Python and Java – with the help of \cite{JYTHON} – share their pools of language bindings, vastly enriching both ecosystems.

While Jython already enables Python code to access Java frameworks and also native JNI-based (Java Native Interface) C extensions, it currently locks out all extensions that use CPython's native API \cite{C-API}. Remember that this does not only affect the actual C extensions, but also all Python frameworks that have a – maybe single, subtle – dependency on such an extension. Dependencies can include:%
\begin{itemize}

\item 

Libraries like NumPy that are written directly in terms of the C-API. These libraries, which in turn link native libraries like BLAS, are widely used in the Python ecosystem, especially in scientific code.
\item 

\cite{CYTHON} is a popular tool to build optimized C code from Python source that has been annotated with types and other declaration, using the C-API to link.
\item 

The \cite{CTYPES} and \cite{CFFI} modules, comparable to \cite{JNA} and \cite{JNR} in the Java-world respectively, are other popular means of providing support for C bindings, also all written to use the C-API.
\item 

\cite{SWIG}, \cite{PYREX} (from which Cython was derived), Boost.Python (\cite{BOOSTPY}) and \cite{SIP} are further tools that create extensions using the C-API.
\end{itemize}

\cite{JyNI} (Jython Native Interface) is going to improve this situation. It is a compatibility layer that implements CPython's C-API on top of JNI and Jython. This way it enables Jython to load native CPython extensions and use them the same way as one would do in CPython. To leverage this functionality, no modification to Python code or C extension source code is required – one just needs to add \texttt{JyNI.jar} to Jython's classpath (along with its binary libraries). That means JyNI is binary compatible with existing builds of CPython extensions.

Developing JyNI is no trivial task, neither is it completed yet. Main reason for this is Python's rather complex C-API that allows to access internal structures, methods and memory positions directly or via C preprocessor macros (in some sense CPython simply exposes its own internal API via a set of public headers). Existing extensions frequently \emph{do} make use of this, so it is not a purely academical concern. Concepts like Python's global interpreter lock (GIL), exception handling and the buffer protocol are further aspects that complicate writing JyNI. \cite{PMB_PL15} mentions the same issues from \cite{PyPy}'s perspective and confirms the difficulty of providing CPython's native API.

By far the most complex issue overall – and main focus of this work – is garbage collection. Unlike JNI, CPython offers C-level access to its garbage collector (GC) and extensions can use it to manage their memory. Note that in contrast to Java's mark-and-sweep-based GC, CPython's GC uses reference-counting and performs reference-cycle-search. Adopting the original CPython-GC for native extensions is no feasible solution in JyNI-context as pure Java-objects can become part of reference-cycles that would be untraceable and cause immortal trash. Section \DUrole{ref}{why-is-garbage-collection-an-issue} describes this issue in detail.

While there are conceptual solutions for all mentioned issues, JyNI does not yet implement the complete C-API and currently just works as a proof of concept. It can already load the ctypes extension and perform some tasks with it; e.g. it can successfully run several tests of a slightly modified PyOpenGL (\cite{Py_OGL}) version (some very long methods had to be split to comply with limited method-length supported by the JVM, see \cite{Jy_OGL}).
We are working to complete ctypes support as soon as possible, since many Python libraries, e.g. for graphics (PyOpenGL) and 3D-plotting etc., have only a single native dependency on it.
NumPy and SciPy are other important extensions we plan to support with priority. As of version alpha.4 JyNI can import the latest NumPy repository version (upcoming NumPy 1.12) and perform some very basic operations with it (cf. section \DUrole{ref}{experimental-numpy-import}); however a full support is still in unknown distance (cf. section \DUrole{ref}{roadmap}).

\subsection{Overview%
  \label{overview}%
}

JyNI's basic functionality has been described in \cite{JyNI_EP13}. After giving a short comprehension in section \DUrole{ref}{implementation} we will focus on garbage collection in section \DUrole{ref}{garbage-collection}. For usage examples and a demonstration-guide also see \cite{JyNI_EP13}.
Section \DUrole{ref}{weak-references} focuses on the difficulties regarding weak references and section \DUrole{ref}{examples} discusses some demonstration examples.

\subsection{Related Work%
  \label{related-work}%
}

There have been similar efforts in other contexts.%
\begin{itemize}

\item 

\cite{JEP} and \cite{JPY} can bridge Java and Python by embedding the CPython interpreter. However, none of
these approaches aims for integration with Jython. In contrast to that, JyNI is entirely based on
Jython and its runtime.
\item 

Ironclad (\cite{ICLD}) is a JyNI-equivalent approach for IronPython (\cite{IRPY}).
\item 

PyMetabiosis (\cite{PMB}) provides C extension support in PyPy to some extent by embedding the CPython
interpreter. So its approach is comparable to \cite{JEP} and \cite{JPY}.
\item 

\cite{CPYEXT} refers to \cite{PyPy}'s in-house (incomplete) C extension API support. The approach differs from
\cite{JyNI} by requiring recompilation and sometimes adjustments of the extensions using PyPy-specific
headers\DUfootnotemark{id30}{notecpyext}{1}.
\end{itemize}
\DUfootnotetext{notecpyext}{id30}{1}{\phantomsection\label{notecpyext}
This yields advantages and disadvantages compared to JyNI; discussing these is out of scope for this work.}

None of the named approaches reached a sufficient level of functionality/compatibility, at least not for current language versions (some of them used to work to some extend, but became unmaintained). In the Python ecosystem the C extension API has been an ongoing issue since its beginning. PyPy famously has been encouraging developers to favor CFFI over C extension API, as it is the only existing approach that has been designed to be well portable to other Python implementations. However, even if this effort would work out, there would be so many legacy extensions around that a serious move to CFFI won't be done in foreseeable future\DUfootnotemark{id31}{notecffi}{2}.%
\DUfootnotetext{notecffi}{id31}{2}{\phantomsection\label{notecffi}
Our plan is to support the CPython-variant of CFFI in JyNI as an alternative to the ideal approach of a direct port. Creating a true Jython version of CFFI would be a distinct project and was partly done based on \cite{JNR}/JFFI.}

For some of these projects JyNI's GC-approach might be a relevant inspiration, as they face the same problem if it comes to native extensions. There are even vague considerations for CPython to switch to mark-and-sweep-based GC one day to enable a GIL-free version (c.f. \cite{PY3_PL15}). Background here is the fact that reference-counting-based garbage collection is the main reason why CPython needs a GIL: Current reference counters are not atomic and switching to atomic reference counters yields insufficient performance.
In context of a mark-and-sweep-based garbage collection in a future CPython the JyNI GC-approach could potentially be adopted to support legacy extensions and provide a smooth migration path.

\section{Implementation%
  \label{implementation}%
}

In order to bridge Jython's and CPython's concepts of PyObjects, we apply three
different techniques, depending on the PyObject's implementation details.\begin{figure}[h]\noindent\makebox[\columnwidth][c]{\includegraphics[scale=0.26]{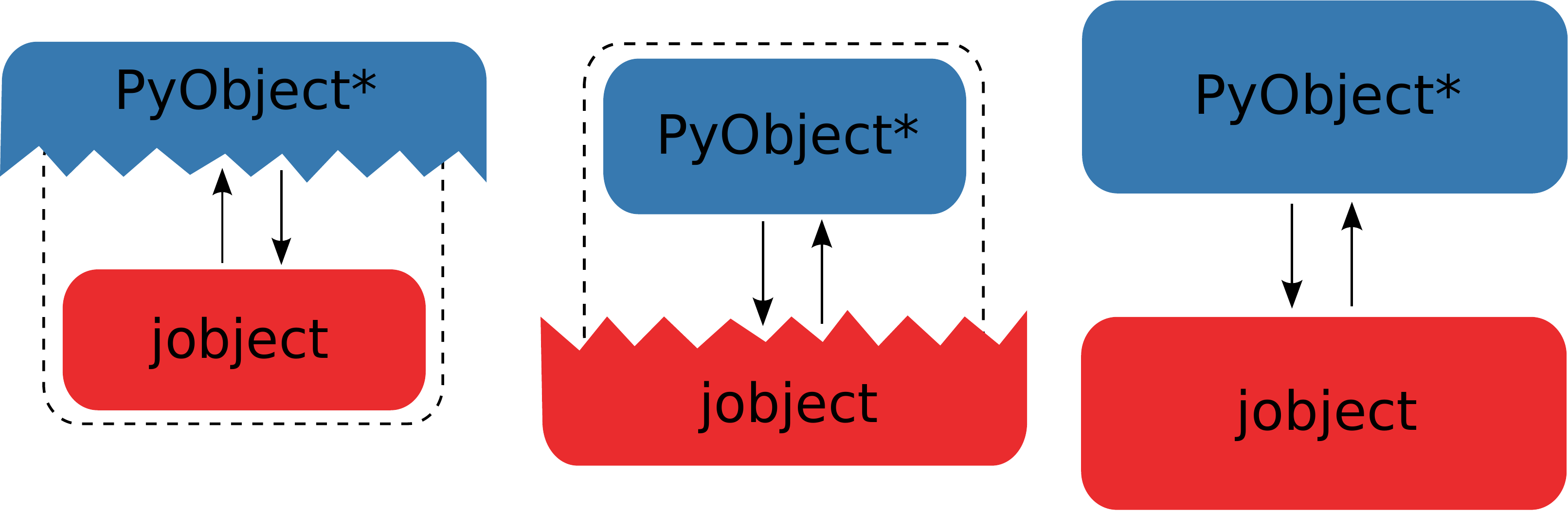}}
\caption{Approaches to bridge PyObjects. \emph{Left}: Native PyObject wraps Java. \emph{Center}: Java-PyObject wraps native one. \emph{Right}: Objects are mirrored. \DUrole{label}{modi}}
\end{figure}

The basic approach is to back the C-API of PyObject by a Java-PyObject via JNI.
This would avoid data synchronization issues, but is only feasible if there are matching counterparts of the PyObject type in Jython and CPython (fig. \DUrole{ref}{modi}, left).
For CPython-specific types we can do it the other way round  (fig. \DUrole{ref}{modi}, center). Another problem is that CPython API defines macros in public headers that access PyObjects' internal data. To deal with these, we sometimes have to mirror the object (fig. \DUrole{ref}{modi}, right).
This might involve data synchronization issues, but luckily macros mostly exist for immutable types, so initial synchronization is sufficient. \cite{JyNI_EP13} describes this in more detail.

\subsection{Global interpreter lock%
  \label{global-interpreter-lock}%
}

As mentioned before, CPython needs a GIL, because its reference-counting-based GC uses non-atomic reference counters. That means that CPython is entirely single-threaded in its usual operation mode.
A native extension can explicitly release the GIL by inserting the macros \texttt{Py\_BEGIN\_ALLOW\_THREADS} and \texttt{Py\_END\_ALLOW\_THREADS} to deal with multiple threads and related things like input events (e.g. Tkinter needs this). In the potentially multithreaded code between these macros it is the extension's own responsibility to refrain from non-thread-safe operations like incrementing or decrementing reference counters. This can be error-prone and challenging as the extension must ensure this also for eventually called external methods.

Jython on the other hand has no GIL and is fully multithreaded based on Java's threading architecture. This does not mean multithreading would be trivial – one still has to care for concurrency issues and thread synchronization, but the whole machinery Java came up with for this topic is available to deal with it.

From JyNI's perspective this is a difficult situation. On one hand we want to avoid regressions on Jython-site, especially regarding an important feature like GIL-freeness. On the other hand, native C extensions might rely on CPython's GIL.
So as a compromise JyNI provides a GIL for native site that is acquired by any thread that enters native code. On returning to Java code, i.e. finishing the native method call, the JyNI-GIL is released. Note that re-entering Java-site by doing a Java call from a native method would \emph{not} release the GIL. In case it is desired to release the GIL for such a re-entering of Java-site or in some other situation, JyNI also supports \texttt{Py\_BEGIN\_ALLOW\_THREADS} and \texttt{Py\_END\_ALLOW\_THREADS} from CPython. This architecture implies that multiple threads can exist on Java-site, while only one thread can exist on native site at the same time (unless allow-threads macros are used). When combining multithreaded Jython code with JyNI it is the developer's responsibility to avoid issues that might arise from this design.

\section{Garbage Collection%
  \label{garbage-collection}%
}

While there are standard approaches for memory management in context of JNI,
none of these is applicable to JyNI. In this section we sketch the default
approaches, illustrate why they fail and finally provide a feasible solution.

\subsection{Why is Garbage Collection an issue?%
  \label{why-is-garbage-collection-an-issue}%
}

Consider a typical JNI-scenario where a native object is accessed from Java.
Usually one would have a Java-object (a “peer”) that stores the native
memory address of the C-object (i.e. a pointer to it) in a \texttt{long}-field. The
naive approach to do memory management would be a \texttt{finalize}-method
in the peer-class. This finalizer would then trigger a native \texttt{free}-call
on the stored memory-handle. However, finalizers are considered bad style in
Java as they impact GC-efficiency. The recommended approach for this scenario
is based on weak references and a reference-queue (c.f. \cite{JREF}).\begin{figure}[h]\noindent\makebox[\columnwidth][c]{\includegraphics[scale=0.42]{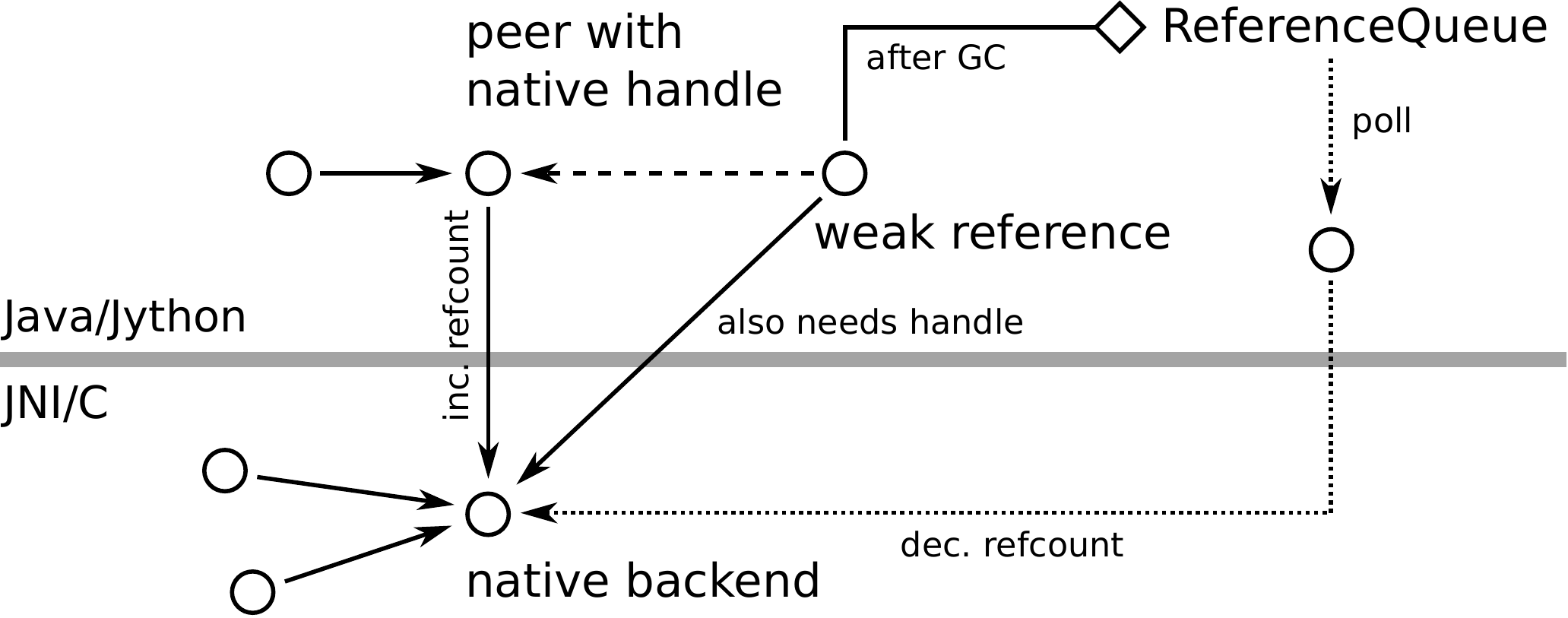}}
\caption{Ordinary JNI memory management \DUrole{label}{oJNImm}}
\end{figure}

Figure \DUrole{ref}{oJNImm} sketches the following procedure:%
\begin{itemize}

\item 

a \texttt{java.lang.ref.WeakReference} is used to track the peer
\item 

actually we store a copy of the peer's native handle in a subclass of \texttt{java.lang.ref.WeakReference}
\item 

a \texttt{java.lang.ref.ReferenceQueue} is registered with the weak reference
\item 

after every run, Java-GC automatically adds cleared weak references to such
a queue if one is registered
(this is Java's variant of Python's weak reference callbacks)
\item 

we poll from the reference queue and clean up the corresponding native resource
\item 

since other native objects might need the resource, we don't call \texttt{free},
but instead perform reference counting
\end{itemize}

So far this would work, but JyNI also needs the opposite scenario with
a native peer backed by a Java-object (figure \DUrole{ref}{nnJ0}).\begin{figure}[h]\noindent\makebox[\columnwidth][c]{\includegraphics[scale=0.42]{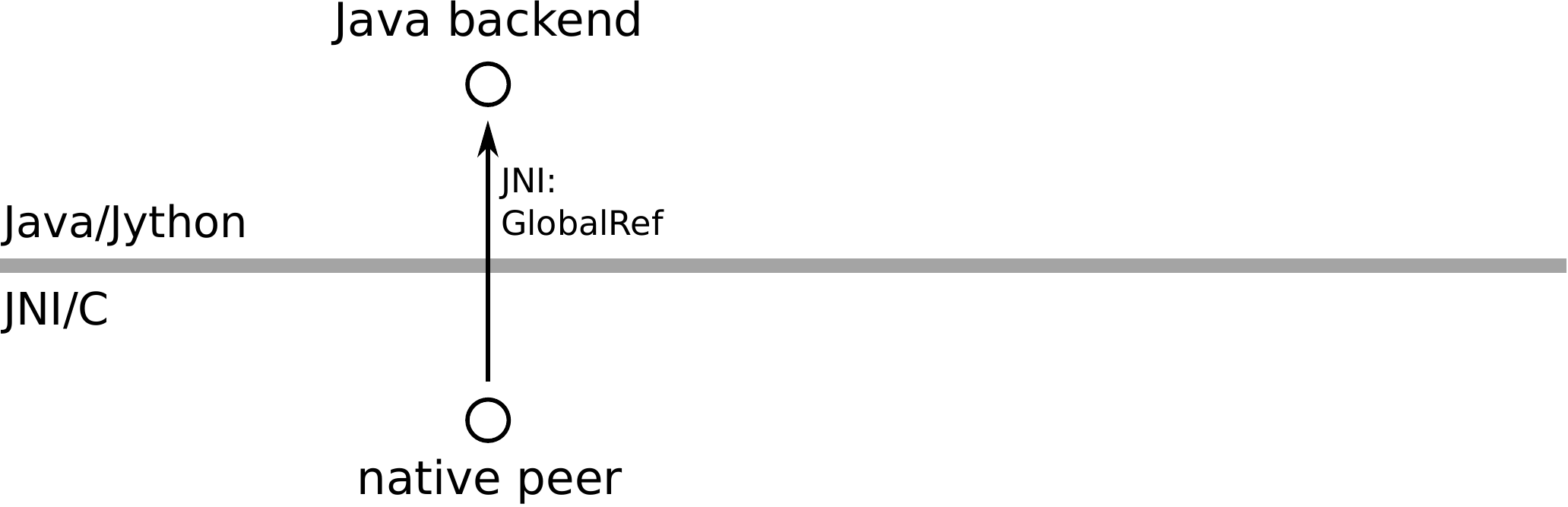}}
\caption{A native peer backed by a Java-object \DUrole{label}{nnJ0}}
\end{figure}

To prevent Java-GC from destroying the Java-backend while it is in use, JNI offers
the concept of global references – JNI-\texttt{GlobalRef}-objects. However, native code
must explicitly create and release such global references. During the lifetime of a
native global reference the Java-site referent is immortal. Now consider the referent
would hold further references to other Java-objects. The reference chain could at
some point include an object that is a peer like shown in figure \DUrole{ref}{oJNImm}. This peer
would keep alive a native object by holding a reference-increment on it. If
the native object also holds reference-increments of other native objects this
can create a pathological reference cycle like illustrated in figure \DUrole{ref}{aprc}.\begin{figure}[h]\noindent\makebox[\columnwidth][c]{\includegraphics[scale=0.42]{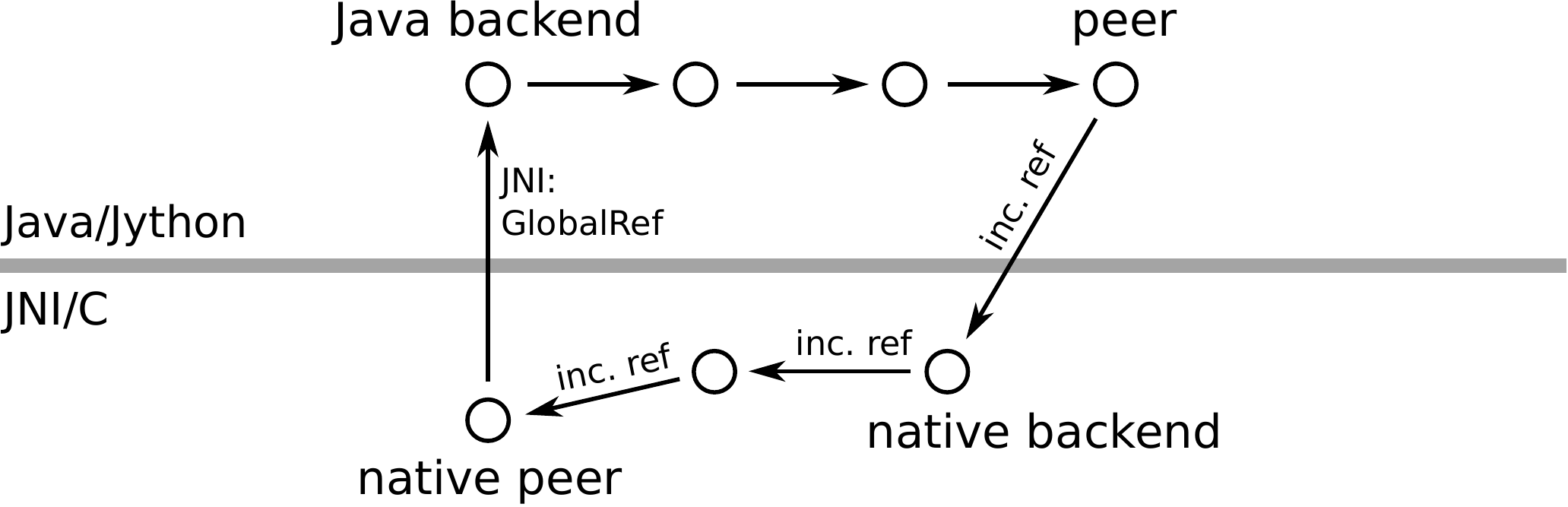}}
\caption{A pathological reference cycle \DUrole{label}{aprc}}
\end{figure}

This kind of cycle cannot be cleared by Java-GC as the \texttt{GlobalRef} prevents it.
Native reference cycle search like known from CPython could not resolve the cycle
either, because it cannot be traced through Java-site. For debugging purposes we actually
added a traverseproc-mechanism to Jython that would allow to trace references
through Java-site, but to clear such a cycle in general just tracing Java-site
references is not sufficient; Java-site reference counting would be required. This
in turn would Jython require to have a GIL, which would be an unacceptable regression.

\subsection{How JyNI solves it (basic approach)%
  \label{how-jyni-solves-it-basic-approach}%
}

To solve this issue, JyNI explores the native reference graph using CPython's traverseproc
mechanism. This is a mechanism PyObjects must implement in order to be traceable by
CPython's garbage collector, i.e. by the code that searches for reference cycles. Basically
a \texttt{PyObject} exposes its references to other objects this way. While JyNI explores the native
reference graph, it mirrors it on Java-site using some minimalistic head-objects
(\texttt{JyNIGCHead} s); see figure \DUrole{ref}{rnrg}. Note that with this design, also Java-objects,
especially Jython-PyObjects can participate in the reference graph and keep parts of it alive.
The kind of object that needed a JNI-\texttt{GlobalRef} in figure \DUrole{ref}{aprc}, can now be tracked by a JNI-\texttt{WeakGlobalRef} while it is kept alive by the mirrored reference graph on Java-site as figure \DUrole{ref}{rnrg} illustrates.
\begin{figure}[H]\noindent\makebox[\columnwidth][c]{\includegraphics[scale=0.42]{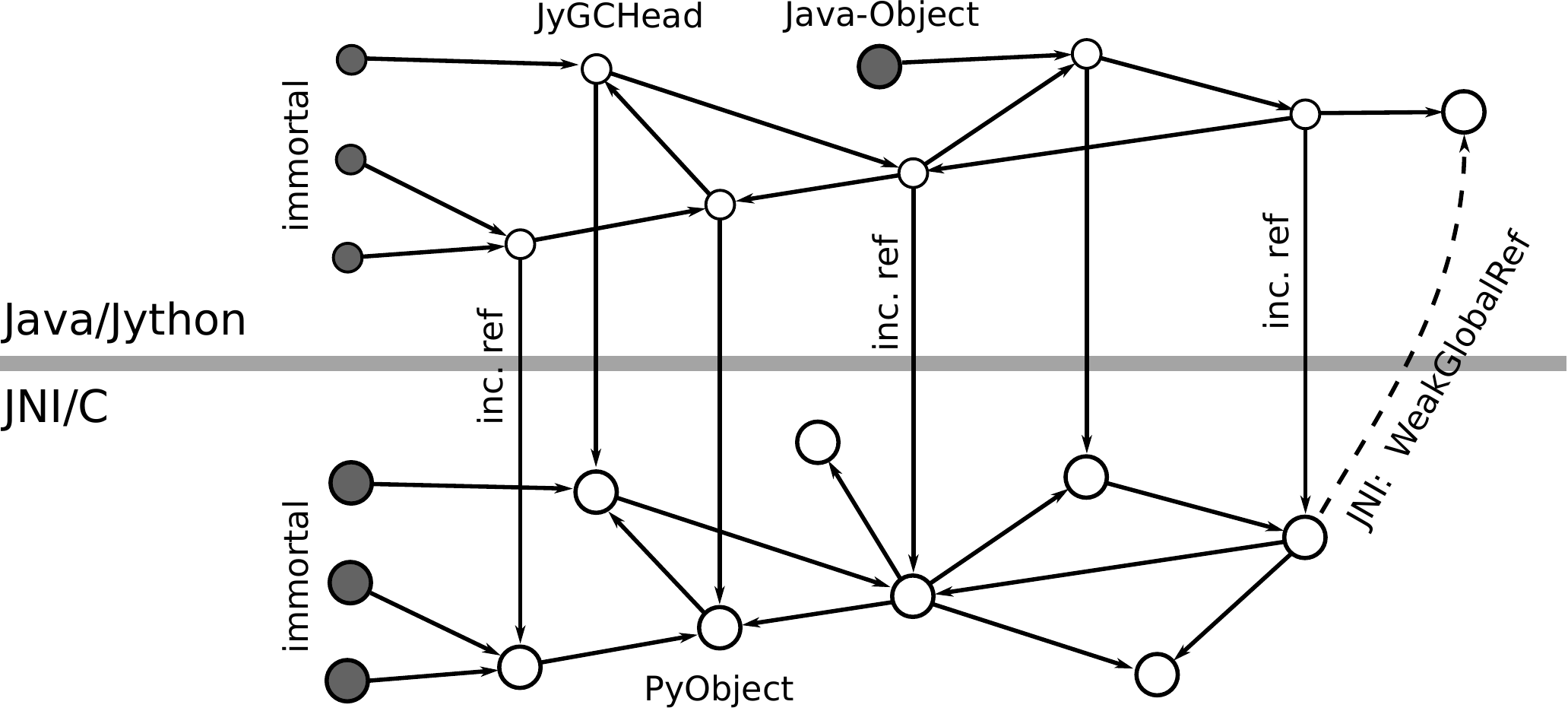}}
\caption{reflected native reference graph \DUrole{label}{rnrg}}
\end{figure}

If a part of the (native) reference-graph becomes unreachable (figure \DUrole{ref}{cuo}), this is
reflected (asynchronously) on Java-site. On its next run, Java-GC will collect this
subgraph, causing weak references to detect deleted objects and then release native references.\begin{figure}[h]\noindent\makebox[\columnwidth][c]{\includegraphics[scale=0.42]{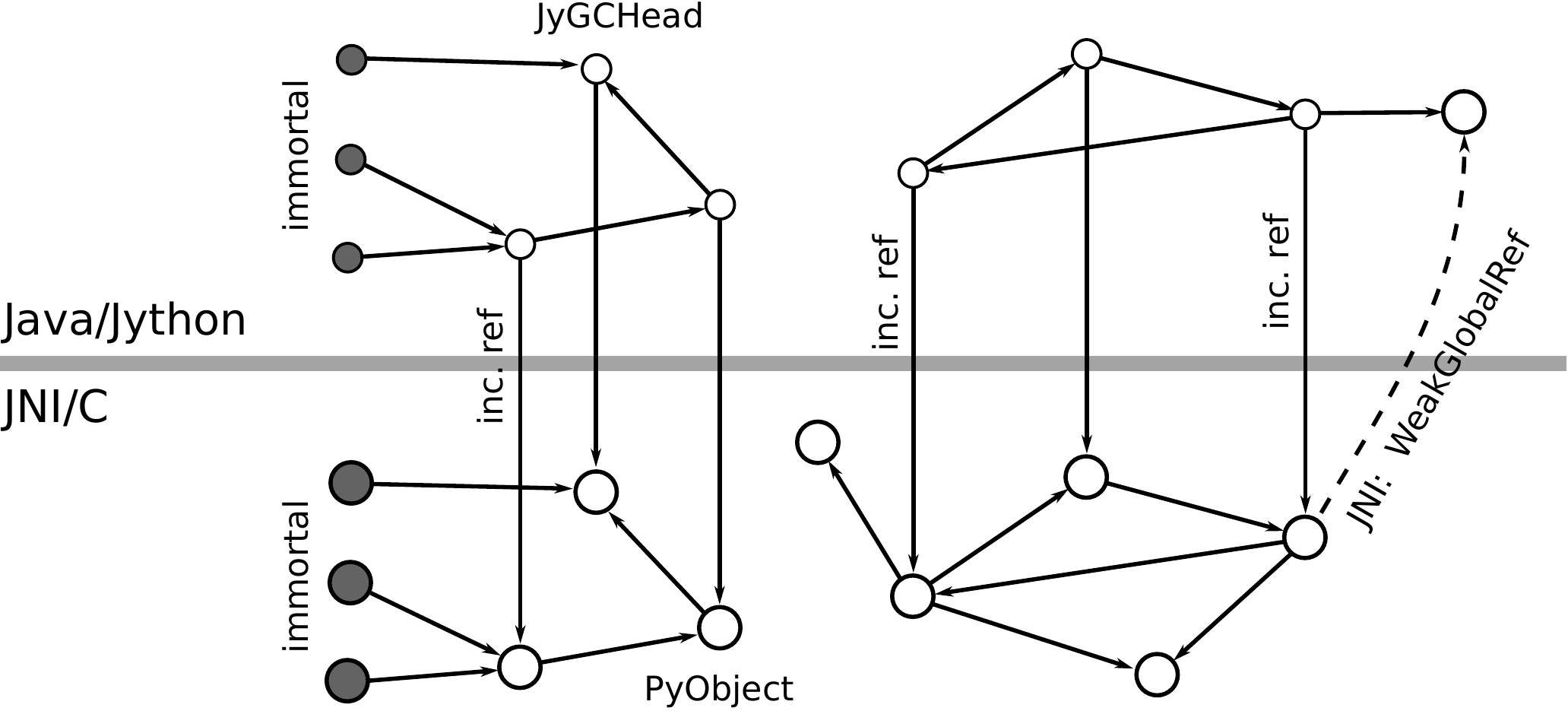}}
\caption{clearing unreachable objects \DUrole{label}{cuo}}
\end{figure}

\subsection{How JyNI solves it (hard case)%
  \label{how-jyni-solves-it-hard-case}%
}

The fact that the reference-graph is mirrored asynchronously can lead to bad situations.
While JyNI features API that allows C code to report changes of the graph, we cannot
enforce third-party-written native extensions to report such changes. However, we made
sure that all built-in types instantaneously send updates to Java-site on modification.

Now consider that a native extension changes the reference graph silently (e.g. using macro
access to a PyObject) and Java's GC
runs before this change was mirrored to Java-site. In that case two types of errors could
normally happen:\newcounter{listcnt0}
\begin{list}{\arabic{listcnt0})}
{
\usecounter{listcnt0}
\setlength{\rightmargin}{\leftmargin}
}

\item 

Objects might be deleted that are still in use
\item 

Objects that are not in use any more persist\end{list}

The design applied in JyNI makes sure that only the second type of error can happen and this only
temporarily, i.e. objects might persist for an additional GC-cycle or two, but not forever.
To make sure that the first kind of error cannot happen, we check a to-be-deleted native
reference subgraph for inner consistency before actually deleting it.\begin{figure}[h]\noindent\makebox[\columnwidth][c]{\includegraphics[scale=0.42]{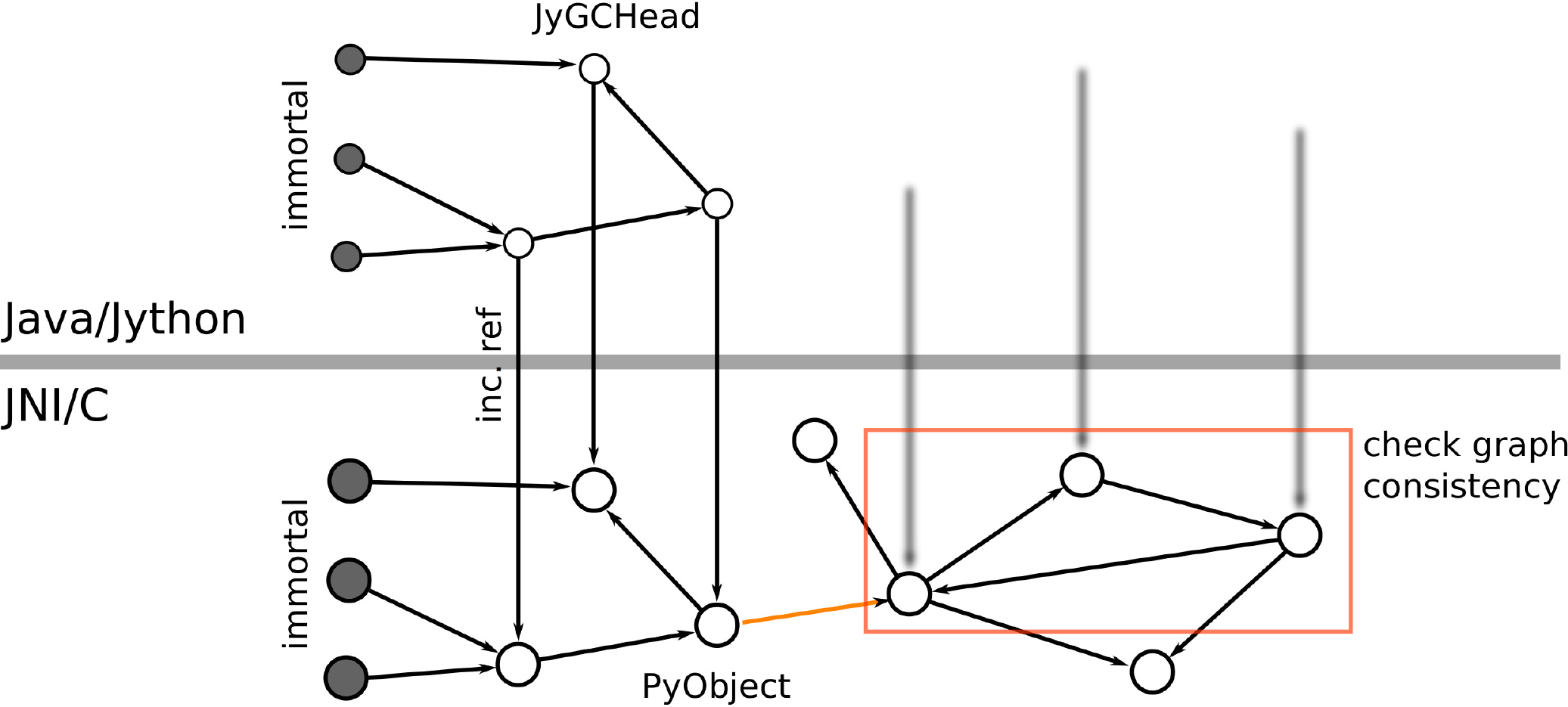}}
\caption{graph must be checked for inner consistency (GC ran before orange connection was mirrored to Java-site) \DUrole{label}{constcy}}
\end{figure}

If not all native reference counts are explainable within this subgraph
(c.f. figure \DUrole{ref}{constcy}), we redo the exploration of participating
PyObjects and update the mirrored graph on Java-site.\begin{figure}[h]\noindent\makebox[\columnwidth][c]{\includegraphics[scale=0.42]{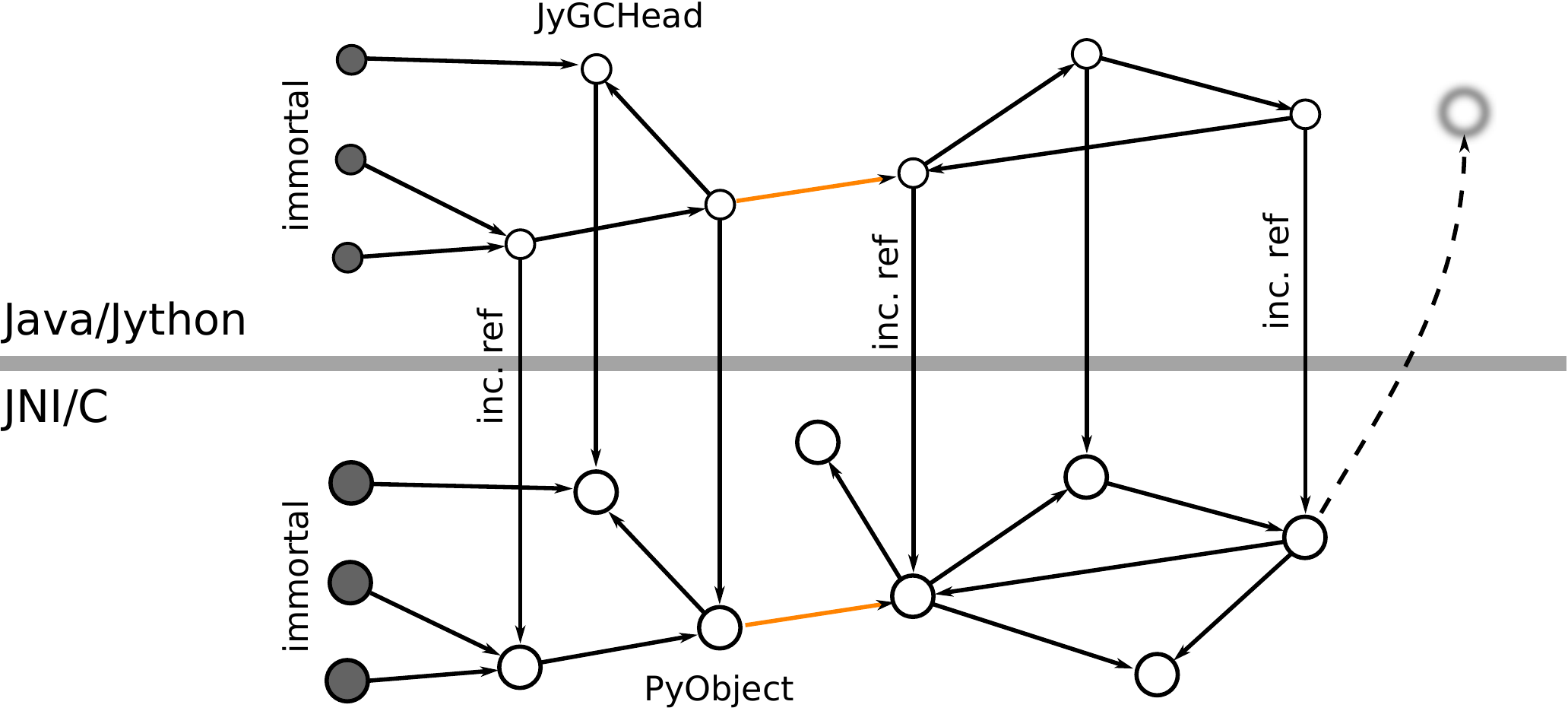}}
\caption{recreated graph \DUrole{label}{recreated}}
\end{figure}

While we can easily recreate the GC-heads, there might be PyObjects that
were weakly reachable from native site and were swept by Java-GC. In order
to restore such objects, we must perform a resurrection
(c.f. figure \DUrole{ref}{resurrected}).\begin{figure}[h]\noindent\makebox[\columnwidth][c]{\includegraphics[scale=0.42]{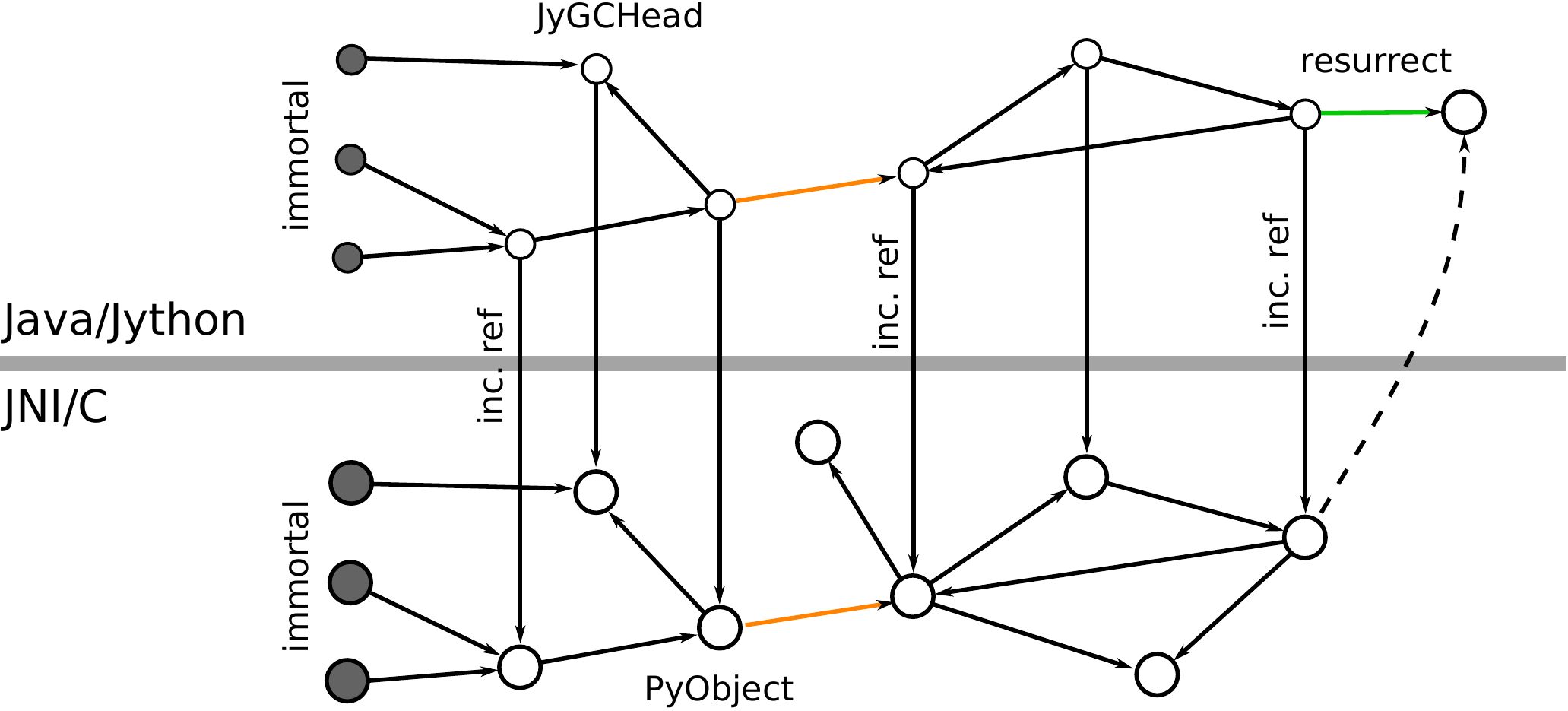}}
\caption{resurrected Java-backend \DUrole{label}{resurrected}}
\end{figure}

The term “object-resurrection” refers to a situation where an object was
garbage-collected, but has a finalizer that restores a strong reference
to it. Note that while resurrection is not recommended – actually the
possibility of a resurrection is the main reason why finalizers are
not recommended – it is a legal operation. So certain GC-heads need to be able
to resurrect an underlying Jython-PyObject and thus must have a finalizer.
Since only certain objects can be subject to a silent reference-graph
modification, it is sufficient to let only GC-heads attached to these objects
implement finalizers – we use finalizers only where really needed.

\subsection{Testing native garbage collection%
  \label{testing-native-garbage-collection}%
}

Since the proposed garbage collection algorithm is rather involved, it is
crucial to have a good way to test it. To achieve this we developed a
monitoring concept that is capable of tracking native allocations, finalizations,
re- and deallocations. The class \texttt{JyNI.JyReferenceMonitor} can – if native
monitoring is enabled – list at any time all natively allocated objects,
their reference counts, timestamps for allocation, finalization, re-
and deallocations and the corresponding code positions (file and line-number)
that performed the memory operations. Unless explicitly cleared, it can also
provide history of these actions. The method \texttt{listLeaks()} lists all currently
allocated native objects (actually these are not necessarily leaks, if the method
is not called at the end of a program or test). While \texttt{listLeaks()} is useful for
debugging, \texttt{getCurrentNativeLeaks()} provides a list that is ideal for unit
testing. E.g. one can assert that no objects are leaked:\begin{Verbatim}[commandchars=\\\{\},fontsize=\footnotesize]
\PY{k+kn}{from} \PY{n+nn}{JyNI} \PY{k+kn}{import} \PY{n}{JyReferenceMonitor} \PY{k}{as} \PY{n}{monitor}
\PY{c+c1}{\PYZsh{}...}
\PY{n+nb+bp}{self}\PY{o}{.}\PY{n}{assertEqual}\PY{p}{(}
    \PY{n+nb}{len}\PY{p}{(}\PY{n}{monitor}\PY{o}{.}\PY{n}{getCurrentNativeLeaks}\PY{p}{(}\PY{p}{)}\PY{p}{)}\PY{p}{,} \PY{l+m+mi}{0}\PY{p}{)}
\end{Verbatim}
The native counterpart of \texttt{JyNI.JyReferenceMonitor} is \texttt{JyRefMonitor.c}.
Its header defines the \texttt{JyNIDebug} macro family, wich we insert into C code
wherever memory operations occur (mainly in \texttt{obmalloc.c} and various inlined
allocations in \texttt{stringobject.c}, \texttt{intobject.c} etc.).

Consider the following demonstration code:\begin{Verbatim}[commandchars=\\\{\},fontsize=\footnotesize]
\PY{k+kn}{import} \PY{n+nn}{time}
\PY{k+kn}{from} \PY{n+nn}{java.lang} \PY{k+kn}{import} \PY{n}{System}
\PY{k+kn}{from} \PY{n+nn}{JyNI} \PY{k+kn}{import} \PY{n}{JyReferenceMonitor} \PY{k}{as} \PY{n}{monitor}
\PY{k+kn}{import} \PY{n+nn}{DemoExtension}
\PY{n}{JyNI}\PY{o}{.}\PY{n}{JyRefMonitor\PYZus{}setMemDebugFlags}\PY{p}{(}\PY{l+m+mi}{1}\PY{p}{)}
\PY{n}{lst} \PY{o}{=} \PY{p}{(}\PY{p}{[}\PY{l+m+mi}{0}\PY{p}{,} \PY{l+s+s2}{\PYZdq{}}\PY{l+s+s2}{test}\PY{l+s+s2}{\PYZdq{}}\PY{p}{]}\PY{p}{,}\PY{p}{)}
\PY{n}{l}\PY{p}{[}\PY{l+m+mi}{0}\PY{p}{]}\PY{p}{[}\PY{l+m+mi}{0}\PY{p}{]} \PY{o}{=} \PY{n}{lst}
\PY{n}{DemoExtension}\PY{o}{.}\PY{n}{argCountToString}\PY{p}{(}\PY{n}{lst}\PY{p}{)}
\PY{k}{del} \PY{n}{lst}
\PY{k}{print} \PY{l+s+s2}{\PYZdq{}}\PY{l+s+s2}{Leaks before GC:}\PY{l+s+s2}{\PYZdq{}}
\PY{n}{monitor}\PY{o}{.}\PY{n}{listLeaks}\PY{p}{(}\PY{p}{)}
\PY{n}{System}\PY{o}{.}\PY{n}{gc}\PY{p}{(}\PY{p}{)}
\PY{n}{time}\PY{o}{.}\PY{n}{sleep}\PY{p}{(}\PY{l+m+mi}{2}\PY{p}{)}
\PY{k}{print} \PY{l+s+s2}{\PYZdq{}}\PY{l+s+s2}{Leaks after GC:}\PY{l+s+s2}{\PYZdq{}}
\PY{n}{monitor}\PY{o}{.}\PY{n}{listLeaks}\PY{p}{(}\PY{p}{)}
\end{Verbatim}
It creates a reference cycle, passes it to a native function and deletes it
afterwards. By passing it to native code, a native counterpart of \texttt{lst} was
created, which cannot be cleared without some garbage collection (also in
CPython it would need the reference cycle searching GC).
We list the leaks before calling Java's GC and after running it.
The output is as follows:%
\begin{quote}\begin{verbatim}
Leaks before GC:
Current native leaks:
140640457447208_GC (list) #2:
    "[([...],), 'test']"_j *38
140640457457768_GC (tuple) #1:
    "(([([...],), 'test'],),)"_j *38
140640457461832 (str) #2: "test"_j *38
140640457457856_GC (tuple) #3:
    "([([...],), 'test'],)"_j *38
Leaks after GC:
no leaks recorded
\end{verbatim}

\end{quote}
We can see that it lists some leaks before running Java's GC. Each line
consists of the native memory position, the type (in parentheses), the
current native reference count indicated by \texttt{\#}, a string representation
and the creation time indicated by \texttt{*} in milliseconds after initialization
of the \texttt{JyReferenceMonitor} class. The postfix \texttt{\_GC} means that the object
is subject to garbage collection, i.e. it can hold references to other objects
and thus participate in cycles. Objects without \texttt{\_GC} will be directly freed
when the reference counter drops to zero. The postfix \texttt{\_j} of the string
representation means that it was generated by Jython rather than by native code.
We close this section by discussing the observed reference counts:%
\begin{itemize}

\item 

The list-object has one reference increment from its \texttt{JyGCHead} and the other
from the tuple at the bottom of the output.
\item 

The first-listed tuple is the argument-tuple and only referenced by its \texttt{JyGCHead}.
\item 

The string is referenced by its \texttt{JyGCHead} and the list.
\item 

The tuple at the bottom is referenced by its \texttt{JyGCHead}, by the list and by
the argument-tuple.
\end{itemize}

\section{Weak References%
  \label{weak-references}%
}

Supporting the \texttt{PyWeakRef} built-in type in JyNI is not as complicated as
garbage collection, but still a notably involved task. This is mainly due
to consistency requirements that are not trivial to fulfill.%
\begin{itemize}

\item 

If a Jython weakref-object is handed to native site, this shall convert
to a CPython weakref-object and vice versa.
\item 

If native code evaluates a native weakref, it shall return exactly the same
referent-PyObject that would have been created if the Java-pendant (if one exists)
was evaluated and the result was handed to native site; also vice versa.
\item 

If a Jython weak reference is cleared, its native pendant shall be cleared either.
Still, none of them shall be cleared as long as its referent is still alive.
\item 

This implies that even if a Jython referent-PyObject was deleted (can happen in mirror-case)
Jython weakref-objects stay alive as long as the native pendant of the referent is alive.
If evaluated, such a Jython weakref-object retrieves the Jython referent by converting
the native referent.
\item 

An obvious requirement is that this shall of course work without keeping the referents
alive or creating some kind of memory leak. JyNI's delicate GC mechanism
must be taken into account to fulfill the named requirements in this context.
\end{itemize}


In the following, we explain JyNI's solution to this issue. We start by explaining the
weakref-concepts of Jython and CPython, completing this section by describing how JyNI
combines them to a consistent solution.
Note that CPython's weakref-module actually introduces three built-in types:%
\begin{itemize}

\item 

\texttt{\_PyWeakref\_RefType} (“weakref”)
\item 

\texttt{\_PyWeakref\_ProxyType} (“weakproxy”)
\item 
\begin{DUlineblock}{0em}
\item[] \texttt{\_PyWeakref\_CallableProxyType}
\item[] (“weakcallableproxy”)
\end{DUlineblock}

\end{itemize}

\subsection{Weak References in Jython%
  \label{weak-references-in-jython}%
}

In Jython the package \texttt{org.python.modules.\_weakref} contains the classes that implement
weak reference support.%
\begin{itemize}

\item 

\texttt{ReferenceType} implements the “weakref”-built-in
\item 

\texttt{ProxyType} implements the “weakproxy”-built-in
\item 

\texttt{CallableProxyType} implements the “weakcallableproxy”-built-in
\end{itemize}

All of them extend \texttt{AbstractReference}, which in turn extends
\texttt{PyObject}.\begin{figure}[h]\noindent\makebox[\columnwidth][c]{\includegraphics[scale=0.55]{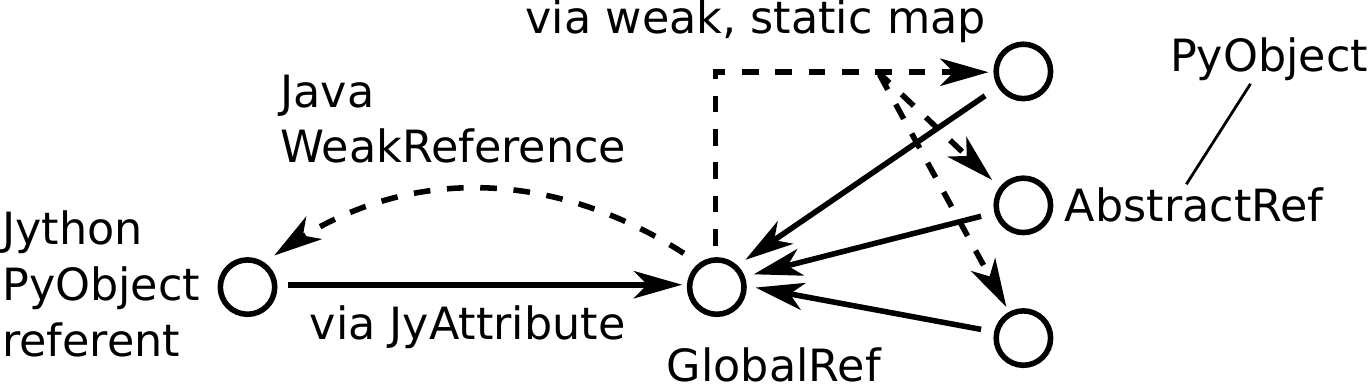}}
\caption{Jython's concept for weak references \DUrole{label}{jythonwr}}
\end{figure}

As figure \DUrole{ref}{jythonwr} illustrates, Jython creates only one Java-style weak reference
per referent. This is created in form of a \texttt{GlobalRef}-object, which extends
\texttt{java.lang.ref.WeakReference}. It stores all Jython weak references pointing to it
in a static, weak-referencing map. This is needed to process potential callbacks when the
reference is cleared. Once created, a \texttt{GlobalRef} is tied to its referent, kept alive
by it and is reused throughout the referent's lifetime. Finally,
\texttt{AbstractReference}-subclasses refer to the \texttt{GlobalRef} corresponding to their actual
referent.

\subsection{Weak References in CPython%
  \label{weak-references-in-cpython}%
}

In CPython, each weakref-type simply contains a reference to its referent without increasing
reference count.\begin{figure}[h]\noindent\makebox[\columnwidth][c]{\includegraphics[scale=0.55]{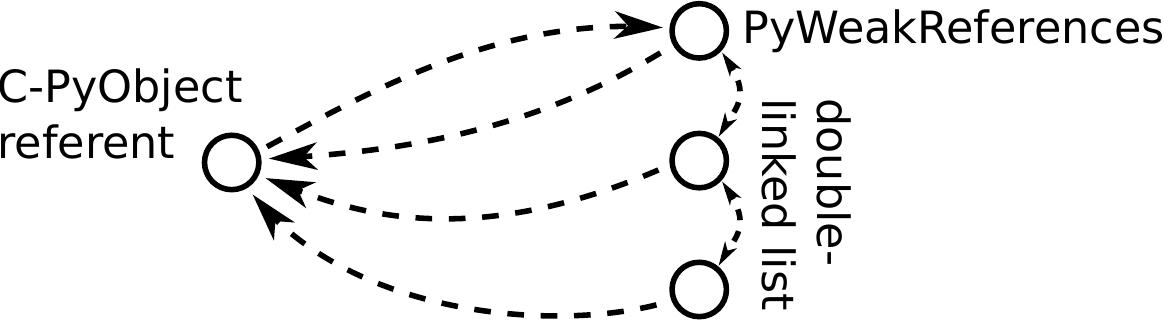}}
\caption{CPython's concept for weak references \DUrole{label}{cpythonwr}}
\end{figure}

Figure \DUrole{ref}{cpythonwr} shows that – like in Jython – referents track
weak references pointing to them; in this case references are stored in a
double-linked list, allowing to iterate them for callback-processing.

\subsection{Weak References in JyNI%
  \label{weak-references-in-jyni}%
}
\begin{figure}[h]\noindent\makebox[\columnwidth][c]{\includegraphics[scale=0.42]{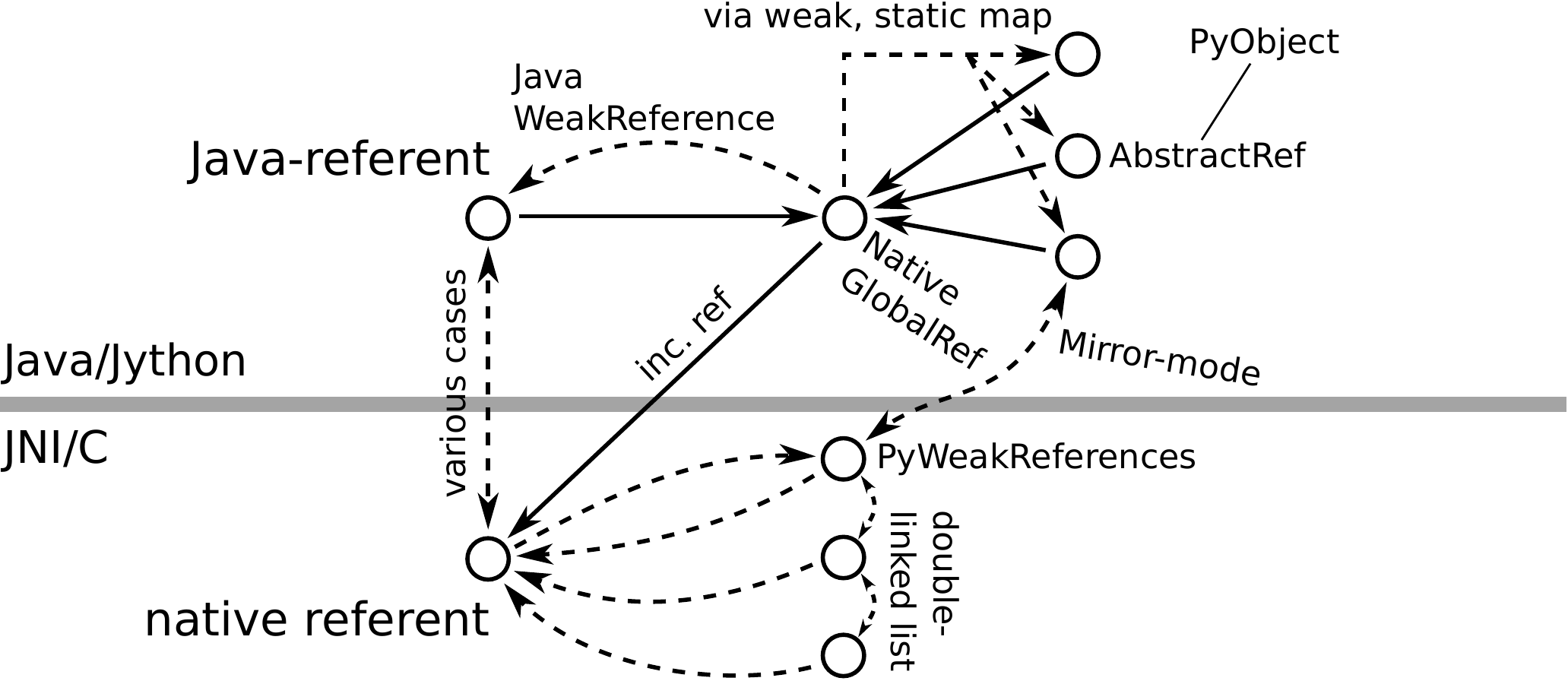}}
\caption{JyNI's concept for weak references \DUrole{label}{jyniwr}}
\end{figure}

JyNI's weak reference support is grounded on CPython's approach on native site and
Jython's approach on Java-site. However, the actual effort is to bridge these approaches
in a consistent way.
To fulfill the requirement for consistent clear-status, we establish a “Java-referent dies
first”-policy. Instead of an ordinary \texttt{GlobalRef}, JyNI uses a subclass called
\texttt{NativeGlobalRef}. This subclass holds a reference-increment for the native counterpart
of its referent. This ensures that the native referent cannot die as long as Jython
weak references exist (see figure \DUrole{ref}{jyniwr}). Otherwise, native weak references might
be cleared earlier than their Jython-pendants. Note that the native ref-increment held by
\texttt{NativeGlobalRef} cannot create a reference-cycle, because it is not reflected by a
\texttt{JyNIGCHead} as seen in figure \DUrole{ref}{rnrg}. Also, the consistency-check shown in figure
\DUrole{ref}{constcy} takes this ref-increment into account, i.e. tracks ref-increments coming from
\texttt{NativeGlobalRef} s separately.\begin{figure}[h]\noindent\makebox[\columnwidth][c]{\includegraphics[scale=0.42]{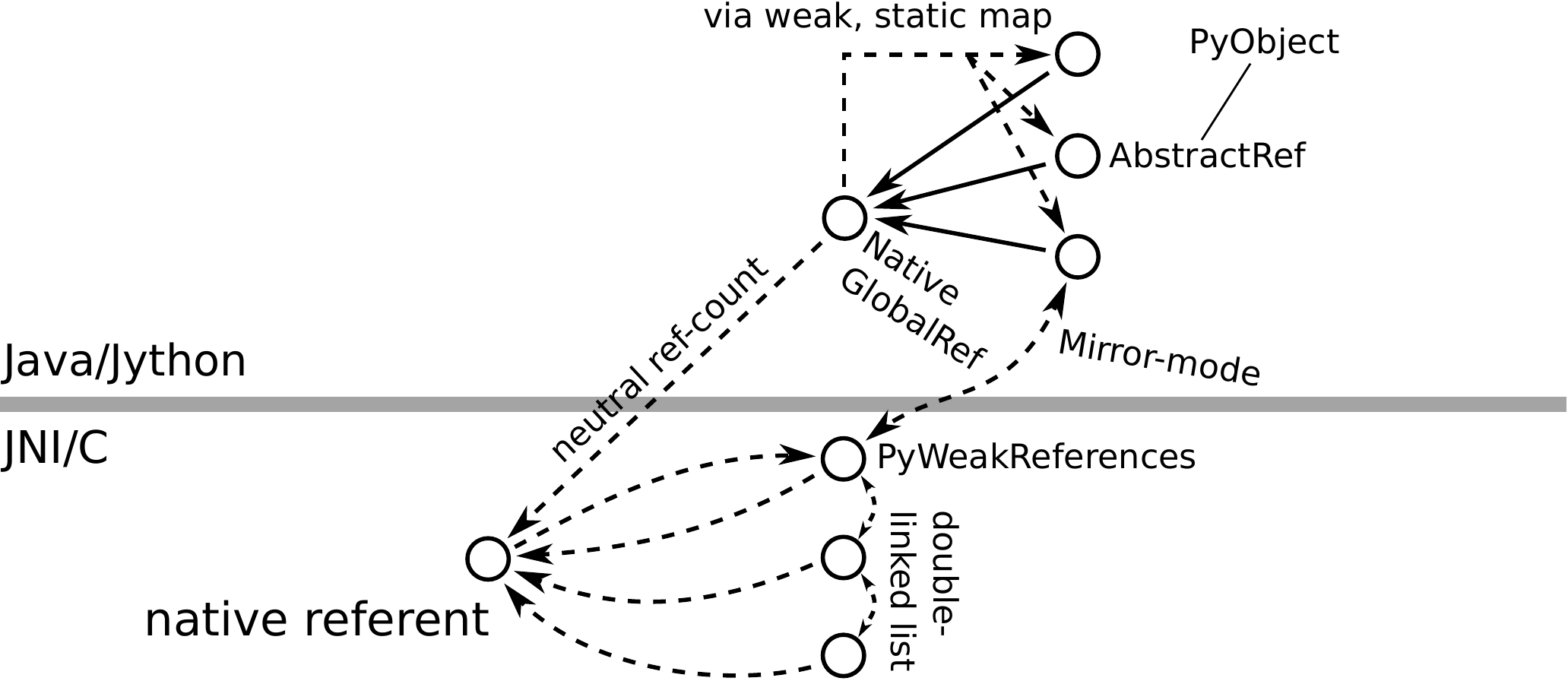}}
\caption{JyNI weak reference after Java-referent was collected \DUrole{label}{jyniwrnj}}
\end{figure}

If the Jython-referent and its native pendant are handled in mirror-mode, it can happen
that the Java-referent is garbage-collected while the native one persists. As soon as the
Jython-referent is collected, its \texttt{NativeGlobalRef} releases the native
reference-increment (see figure \DUrole{ref}{jyniwrnj}). Still, it will not be cleared and process
callbacks, before also the native referent dies. Until then, \texttt{NativeGlobalRef} continues
to be valid – it implements its \texttt{get}-method such that if the Jython-referent is not
available, it is recreated from the native referent. As long as such a retrieved referent is
alive on Java-site, the situation in figure \DUrole{ref}{jyniwr} is restored.

\section{Examples%
  \label{examples}%
}

The code-samples in this section are runnable with Jython 2.7.1 and JyNI 2.7-alpha.3 or newer (alpha.4 for NumPy-import).

\subsection{Using Tkinter from Java%
  \label{using-tkinter-from-java}%
}

In \cite{JyNI_EP13} we demonstrated a minimalistic Tkinter example program that used the original
Tkinter binary bundled with CPython. Here we demonstrate how the same functionality can be
achieved from Java code. This confirms the usability of Python libraries from Java via Jython
and JyNI. While the main magic happens in Jython, it is not completely self-evident that this
is also possible through JyNI and required some internal improvements to work. Remember the
Tkinter-program from \cite{JyNI_EP13}:\begin{Verbatim}[commandchars=\\\{\},fontsize=\footnotesize]
\PY{k+kn}{import} \PY{n+nn}{sys}
\PY{c+c1}{\PYZsh{}Include native Tkinter:}
\PY{n}{sys}\PY{o}{.}\PY{n}{path}\PY{o}{.}\PY{n}{append}\PY{p}{(}\PY{l+s+s1}{\PYZsq{}}\PY{l+s+s1}{/usr/lib/python2.7/lib\PYZhy{}dynload}\PY{l+s+s1}{\PYZsq{}}\PY{p}{)}
\PY{n}{sys}\PY{o}{.}\PY{n}{path}\PY{o}{.}\PY{n}{append}\PY{p}{(}\PY{l+s+s1}{\PYZsq{}}\PY{l+s+s1}{/usr/lib/python2.7/lib\PYZhy{}tk}\PY{l+s+s1}{\PYZsq{}}\PY{p}{)}
\PY{k+kn}{from} \PY{n+nn}{Tkinter} \PY{k+kn}{import} \PY{o}{*}

\PY{n}{root} \PY{o}{=} \PY{n}{Tk}\PY{p}{(}\PY{p}{)}
\PY{n}{txt} \PY{o}{=} \PY{n}{StringVar}\PY{p}{(}\PY{p}{)}
\PY{n}{txt}\PY{o}{.}\PY{n}{set}\PY{p}{(}\PY{l+s+s2}{\PYZdq{}}\PY{l+s+s2}{Hello World!}\PY{l+s+s2}{\PYZdq{}}\PY{p}{)}

\PY{k}{def} \PY{n+nf}{print\PYZus{}text}\PY{p}{(}\PY{p}{)}\PY{p}{:}
    \PY{k}{print} \PY{n}{txt}\PY{o}{.}\PY{n}{get}\PY{p}{(}\PY{p}{)}

\PY{k}{def} \PY{n+nf}{print\PYZus{}time\PYZus{}stamp}\PY{p}{(}\PY{p}{)}\PY{p}{:}
    \PY{k+kn}{from} \PY{n+nn}{java.lang} \PY{k+kn}{import} \PY{n}{System}
    \PY{k}{print} \PY{l+s+s2}{\PYZdq{}}\PY{l+s+s2}{System.currentTimeMillis: }\PY{l+s+s2}{\PYZdq{}}
        \PY{o}{+}\PY{n+nb}{str}\PY{p}{(}\PY{n}{System}\PY{o}{.}\PY{n}{currentTimeMillis}\PY{p}{(}\PY{p}{)}\PY{p}{)}

\PY{n}{Label}\PY{p}{(}\PY{n}{root}\PY{p}{,}
     \PY{n}{text}\PY{o}{=}\PY{l+s+s2}{\PYZdq{}}\PY{l+s+s2}{Welcome to JyNI Tkinter\PYZhy{}Demo!}\PY{l+s+s2}{\PYZdq{}}\PY{p}{)}\PY{o}{.}\PY{n}{pack}\PY{p}{(}\PY{p}{)}
\PY{n}{Entry}\PY{p}{(}\PY{n}{root}\PY{p}{,} \PY{n}{textvariable}\PY{o}{=}\PY{n}{txt}\PY{p}{)}\PY{o}{.}\PY{n}{pack}\PY{p}{(}\PY{p}{)}
\PY{n}{Button}\PY{p}{(}\PY{n}{root}\PY{p}{,} \PY{n}{text}\PY{o}{=}\PY{l+s+s2}{\PYZdq{}}\PY{l+s+s2}{print text}\PY{l+s+s2}{\PYZdq{}}\PY{p}{,}
        \PY{n}{command}\PY{o}{=}\PY{n}{print\PYZus{}text}\PY{p}{)}\PY{o}{.}\PY{n}{pack}\PY{p}{(}\PY{p}{)}
\PY{n}{Button}\PY{p}{(}\PY{n}{root}\PY{p}{,} \PY{n}{text}\PY{o}{=}\PY{l+s+s2}{\PYZdq{}}\PY{l+s+s2}{print timestamp}\PY{l+s+s2}{\PYZdq{}}\PY{p}{,}
        \PY{n}{command}\PY{o}{=}\PY{n}{print\PYZus{}time\PYZus{}stamp}\PY{p}{)}\PY{o}{.}\PY{n}{pack}\PY{p}{(}\PY{p}{)}
\PY{n}{Button}\PY{p}{(}\PY{n}{root}\PY{p}{,} \PY{n}{text}\PY{o}{=}\PY{l+s+s2}{\PYZdq{}}\PY{l+s+s2}{Quit}\PY{l+s+s2}{\PYZdq{}}\PY{p}{,}
        \PY{n}{command}\PY{o}{=}\PY{n}{root}\PY{o}{.}\PY{n}{destroy}\PY{p}{)}\PY{o}{.}\PY{n}{pack}\PY{p}{(}\PY{p}{)}
\PY{n}{root}\PY{o}{.}\PY{n}{mainloop}\PY{p}{(}\PY{p}{)}
\end{Verbatim}
\begin{figure}[h]\noindent\makebox[\columnwidth][c]{\includegraphics[scale=0.36]{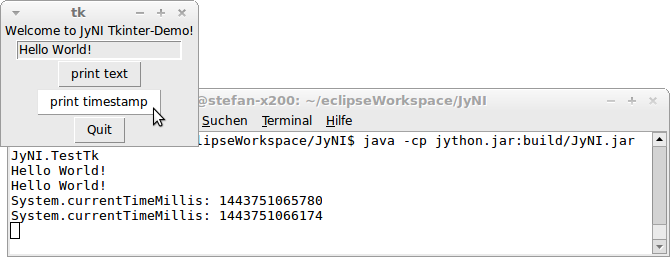}}
\caption{Tkinter demonstration by Java code. Note that the class \texttt{JyNI.TestTk} is executed
rather than \texttt{org.python.util.jython}. \DUrole{label}{tkDemo}}
\end{figure}To translate the program to Java, we must provide type-information via interfaces:\begin{Verbatim}[commandchars=\\\{\},fontsize=\footnotesize]
\PY{k+kn}{import} \PY{n+nn}{org.python.core.*}\PY{o}{;}

\PY{k+kd}{public} \PY{k+kd}{static} \PY{k+kd}{interface} \PY{n+nc}{Label} \PY{o}{\PYZob{}}
    \PY{k+kd}{public} \PY{k+kt}{void} \PY{n+nf}{pack}\PY{o}{(}\PY{o}{)}\PY{o}{;}
\PY{o}{\PYZcb{}}

\PY{k+kd}{public} \PY{k+kd}{static} \PY{k+kd}{interface} \PY{n+nc}{Button} \PY{o}{\PYZob{}}
    \PY{k+kd}{public} \PY{k+kt}{void} \PY{n+nf}{pack}\PY{o}{(}\PY{o}{)}\PY{o}{;}
\PY{o}{\PYZcb{}}

\PY{k+kd}{public} \PY{k+kd}{static} \PY{k+kd}{interface} \PY{n+nc}{Entry} \PY{o}{\PYZob{}}
    \PY{k+kd}{public} \PY{k+kt}{void} \PY{n+nf}{pack}\PY{o}{(}\PY{o}{)}\PY{o}{;}
\PY{o}{\PYZcb{}}

\PY{k+kd}{public} \PY{k+kd}{static} \PY{k+kd}{interface} \PY{n+nc}{Tk} \PY{o}{\PYZob{}}
    \PY{k+kd}{public} \PY{k+kt}{void} \PY{n+nf}{mainloop}\PY{o}{(}\PY{o}{)}\PY{o}{;}
    \PY{k+kd}{public} \PY{k+kt}{void} \PY{n+nf}{destroy}\PY{o}{(}\PY{o}{)}\PY{o}{;}
\PY{o}{\PYZcb{}}

\PY{k+kd}{public} \PY{k+kd}{static} \PY{k+kd}{interface} \PY{n+nc}{StringVar} \PY{o}{\PYZob{}}
    \PY{k+kd}{public} \PY{n}{String} \PY{n+nf}{get}\PY{o}{(}\PY{o}{)}\PY{o}{;}
    \PY{k+kd}{public} \PY{k+kt}{void} \PY{n+nf}{set}\PY{o}{(}\PY{n}{String} \PY{n}{text}\PY{o}{)}\PY{o}{;}
\PY{o}{\PYZcb{}}
\end{Verbatim}
We define the methods backing the button-actions as static methods with a special Python-compliant signature:\begin{Verbatim}[commandchars=\\\{\},fontsize=\footnotesize]
\PY{k+kd}{static} \PY{n}{Tk} \PY{n}{root}\PY{o}{;}
\PY{k+kd}{static} \PY{n}{StringVar} \PY{n}{txt}\PY{o}{;}

\PY{k+kd}{public} \PY{k+kd}{static} \PY{k+kt}{void} \PY{n+nf}{printText}\PY{o}{(}\PY{n}{PyObject}\PY{o}{[}\PY{o}{]} \PY{n}{args}\PY{o}{,}
        \PY{n}{String}\PY{o}{[}\PY{o}{]} \PY{n}{kws}\PY{o}{)} \PY{o}{\PYZob{}}
    \PY{n}{System}\PY{o}{.}\PY{n+na}{out}\PY{o}{.}\PY{n+na}{println}\PY{o}{(}\PY{n}{txt}\PY{o}{.}\PY{n+na}{get}\PY{o}{(}\PY{o}{)}\PY{o}{)}\PY{o}{;}
\PY{o}{\PYZcb{}}

\PY{k+kd}{public} \PY{k+kd}{static} \PY{k+kt}{void} \PY{n+nf}{printTimeStamp}\PY{o}{(}\PY{n}{PyObject}\PY{o}{[}\PY{o}{]} \PY{n}{args}\PY{o}{,}
        \PY{n}{String}\PY{o}{[}\PY{o}{]} \PY{n}{kws}\PY{o}{)} \PY{o}{\PYZob{}}
    \PY{n}{System}\PY{o}{.}\PY{n+na}{out}\PY{o}{.}\PY{n+na}{println}\PY{o}{(}\PY{l+s}{\PYZdq{}System.currentTimeMillis: \PYZdq{}}
            \PY{o}{+} \PY{n}{System}\PY{o}{.}\PY{n+na}{currentTimeMillis}\PY{o}{(}\PY{o}{)}\PY{o}{)}\PY{o}{;}
\PY{o}{\PYZcb{}}

\PY{k+kd}{public} \PY{k+kd}{static} \PY{k+kt}{void} \PY{n+nf}{destroyRoot}\PY{o}{(}\PY{n}{PyObject}\PY{o}{[}\PY{o}{]} \PY{n}{args}\PY{o}{,}
        \PY{n}{String}\PY{o}{[}\PY{o}{]} \PY{n}{kws}\PY{o}{)} \PY{o}{\PYZob{}}
    \PY{n}{root}\PY{o}{.}\PY{n+na}{destroy}\PY{o}{(}\PY{o}{)}\PY{o}{;}
\PY{o}{\PYZcb{}}
\end{Verbatim}
On top of this a rather Java-like main-method can be implemented. Note that constructing objects is still somewhat unhandy, as keywords must be declared in a string-array and explicitly passed to Jython. Calling methods on objects then works like ordinary Java code and is even type-safe based on the declared interfaces.\begin{Verbatim}[commandchars=\\\{\},fontsize=\footnotesize]
\PY{k+kd}{public} \PY{k+kd}{static} \PY{k+kt}{void} \PY{n+nf}{main}\PY{o}{(}\PY{n}{String}\PY{o}{[}\PY{o}{]} \PY{n}{args}\PY{o}{)} \PY{o}{\PYZob{}}
    \PY{n}{PySystemState} \PY{n}{pystate} \PY{o}{=} \PY{n}{Py}\PY{o}{.}\PY{n+na}{getSystemState}\PY{o}{(}\PY{o}{)}\PY{o}{;}
    \PY{n}{pystate}\PY{o}{.}\PY{n+na}{path}\PY{o}{.}\PY{n+na}{add}\PY{o}{(}
            \PY{l+s}{\PYZdq{}/usr/lib/python2.7/lib\PYZhy{}dynload\PYZdq{}}\PY{o}{)}\PY{o}{;}
    \PY{n}{pystate}\PY{o}{.}\PY{n+na}{path}\PY{o}{.}\PY{n+na}{add}\PY{o}{(}\PY{l+s}{\PYZdq{}/usr/lib/python2.7/lib\PYZhy{}tk\PYZdq{}}\PY{o}{)}\PY{o}{;}
    \PY{n}{PyModule} \PY{n}{tkModule} \PY{o}{=} \PY{o}{(}\PY{n}{PyModule}\PY{o}{)}
            \PY{n}{imp}\PY{o}{.}\PY{n+na}{importName}\PY{o}{(}\PY{l+s}{\PYZdq{}Tkinter\PYZdq{}}\PY{o}{,} \PY{k+kc}{true}\PY{o}{)}\PY{o}{;}
    \PY{n}{root} \PY{o}{=} \PY{n}{tkModule}\PY{o}{.}\PY{n+na}{newJ}\PY{o}{(}\PY{n}{Tk}\PY{o}{.}\PY{n+na}{class}\PY{o}{)}\PY{o}{;}
    \PY{n}{txt} \PY{o}{=} \PY{n}{tkModule}\PY{o}{.}\PY{n+na}{newJ}\PY{o}{(}\PY{n}{StringVar}\PY{o}{.}\PY{n+na}{class}\PY{o}{)}\PY{o}{;}
    \PY{n}{txt}\PY{o}{.}\PY{n+na}{set}\PY{o}{(}\PY{l+s}{\PYZdq{}Hello World!\PYZdq{}}\PY{o}{)}\PY{o}{;}

    \PY{n}{Label} \PY{n}{lab} \PY{o}{=} \PY{n}{tkModule}\PY{o}{.}\PY{n+na}{newJ}\PY{o}{(}\PY{n}{Label}\PY{o}{.}\PY{n+na}{class}\PY{o}{,}
            \PY{k}{new} \PY{n}{String}\PY{o}{[}\PY{o}{]}\PY{o}{\PYZob{}}\PY{l+s}{\PYZdq{}text\PYZdq{}}\PY{o}{\PYZcb{}}\PY{o}{,} \PY{n}{root}\PY{o}{,}
            \PY{l+s}{\PYZdq{}Welcome to JyNI Tkinter\PYZhy{}Demo!\PYZdq{}}\PY{o}{)}\PY{o}{;}
    \PY{n}{lab}\PY{o}{.}\PY{n+na}{pack}\PY{o}{(}\PY{o}{)}\PY{o}{;}

    \PY{n}{Entry} \PY{n}{entry} \PY{o}{=} \PY{n}{tkModule}\PY{o}{.}\PY{n+na}{newJ}\PY{o}{(}\PY{n}{Entry}\PY{o}{.}\PY{n+na}{class}\PY{o}{,}
            \PY{k}{new} \PY{n}{String}\PY{o}{[}\PY{o}{]}\PY{o}{\PYZob{}}\PY{l+s}{\PYZdq{}textvariable\PYZdq{}}\PY{o}{\PYZcb{}}\PY{o}{,} \PY{n}{root}\PY{o}{,} \PY{n}{txt}\PY{o}{)}\PY{o}{;}
    \PY{n}{entry}\PY{o}{.}\PY{n+na}{pack}\PY{o}{(}\PY{o}{)}\PY{o}{;}

    \PY{n}{String}\PY{o}{[}\PY{o}{]} \PY{n}{kw\PYZus{}txt\PYZus{}cmd} \PY{o}{=} \PY{o}{\PYZob{}}\PY{l+s}{\PYZdq{}text\PYZdq{}}\PY{o}{,} \PY{l+s}{\PYZdq{}command\PYZdq{}}\PY{o}{\PYZcb{}}\PY{o}{;}
    \PY{n}{Button} \PY{n}{buttonPrint} \PY{o}{=} \PY{n}{tkModule}\PY{o}{.}\PY{n+na}{newJ}\PY{o}{(}\PY{n}{Button}\PY{o}{.}\PY{n+na}{class}\PY{o}{,}
            \PY{n}{kw\PYZus{}txt\PYZus{}cmd}\PY{o}{,} \PY{n}{root}\PY{o}{,} \PY{l+s}{\PYZdq{}print text\PYZdq{}}\PY{o}{,}
            \PY{n}{Py}\PY{o}{.}\PY{n+na}{newJavaFunc}\PY{o}{(}\PY{n}{TestTk}\PY{o}{.}\PY{n+na}{class}\PY{o}{,}
                    \PY{l+s}{\PYZdq{}printText\PYZdq{}}\PY{o}{)}\PY{o}{)}\PY{o}{;}
    \PY{n}{buttonPrint}\PY{o}{.}\PY{n+na}{pack}\PY{o}{(}\PY{o}{)}\PY{o}{;}

    \PY{n}{Button} \PY{n}{buttonTimestamp} \PY{o}{=} \PY{n}{tkModule}\PY{o}{.}\PY{n+na}{newJ}\PY{o}{(}
            \PY{n}{Button}\PY{o}{.}\PY{n+na}{class}\PY{o}{,} \PY{n}{kw\PYZus{}txt\PYZus{}cmd}\PY{o}{,}
            \PY{n}{root}\PY{o}{,} \PY{l+s}{\PYZdq{}print timestamp\PYZdq{}}\PY{o}{,}
            \PY{n}{Py}\PY{o}{.}\PY{n+na}{newJavaFunc}\PY{o}{(}\PY{n}{TestTk}\PY{o}{.}\PY{n+na}{class}\PY{o}{,}
                    \PY{l+s}{\PYZdq{}printTimeStamp\PYZdq{}}\PY{o}{)}\PY{o}{)}\PY{o}{;}
    \PY{n}{buttonTimestamp}\PY{o}{.}\PY{n+na}{pack}\PY{o}{(}\PY{o}{)}\PY{o}{;}

    \PY{n}{Button} \PY{n}{buttonQuit} \PY{o}{=} \PY{n}{tkModule}\PY{o}{.}\PY{n+na}{newJ}\PY{o}{(}\PY{n}{Button}\PY{o}{.}\PY{n+na}{class}\PY{o}{,}
            \PY{n}{kw\PYZus{}txt\PYZus{}cmd}\PY{o}{,} \PY{n}{root}\PY{o}{,} \PY{l+s}{\PYZdq{}Quit\PYZdq{}}\PY{o}{,}
            \PY{n}{Py}\PY{o}{.}\PY{n+na}{newJavaFunc}\PY{o}{(}\PY{n}{TestTk}\PY{o}{.}\PY{n+na}{class}\PY{o}{,}
                    \PY{l+s}{\PYZdq{}destroyRoot\PYZdq{}}\PY{o}{)}\PY{o}{)}\PY{o}{;}
    \PY{n}{buttonQuit}\PY{o}{.}\PY{n+na}{pack}\PY{o}{(}\PY{o}{)}\PY{o}{;}

    \PY{n}{root}\PY{o}{.}\PY{n+na}{mainloop}\PY{o}{(}\PY{o}{)}\PY{o}{;}
\PY{o}{\PYZcb{}}
\end{Verbatim}

\subsection{Using native ctypes%
  \label{using-native-ctypes}%
}
As of version alpha.3 JyNI has experimental support for \cite{CTYPES}. The following code provides a minimalistic example that uses Java- and C-API. Via an std-lib C-call we obtain system time and print it using Java console.\begin{Verbatim}[commandchars=\\\{\},fontsize=\footnotesize]
\PY{k+kn}{import} \PY{n+nn}{sys}
\PY{n}{sys}\PY{o}{.}\PY{n}{path}\PY{o}{.}\PY{n}{append}\PY{p}{(}\PY{l+s+s1}{\PYZsq{}}\PY{l+s+s1}{/usr/lib/python2.7/lib\PYZhy{}dynload}\PY{l+s+s1}{\PYZsq{}}\PY{p}{)}

\PY{k+kn}{import} \PY{n+nn}{ctypes}
\PY{k+kn}{from} \PY{n+nn}{java.lang} \PY{k+kn}{import} \PY{n}{System}

\PY{n}{libc} \PY{o}{=} \PY{n}{ctypes}\PY{o}{.}\PY{n}{CDLL}\PY{p}{(}\PY{l+s+s1}{\PYZsq{}}\PY{l+s+s1}{libc.so.6}\PY{l+s+s1}{\PYZsq{}}\PY{p}{)}
\PY{k}{print} \PY{n}{libc}
\PY{k}{print} \PY{n}{libc}\PY{o}{.}\PY{n}{time}
\PY{n}{System}\PY{o}{.}\PY{n}{out}\PY{o}{.}\PY{n}{println}\PY{p}{(}\PY{l+s+s1}{\PYZsq{}}\PY{l+s+s1}{Timestamp: }\PY{l+s+s1}{\PYZsq{}}\PY{o}{+}\PY{n+nb}{str}\PY{p}{(}\PY{n}{libc}\PY{o}{.}\PY{n}{time}\PY{p}{(}\PY{l+m+mi}{0}\PY{p}{)}\PY{p}{)}\PY{p}{)}
\end{Verbatim}
The output is as follows:%
\begin{quote}\begin{verbatim}
<CDLL 'libc.so.6', handle 83214548 at 2>
<_FuncPtr object at 0x7f897c7165d8>
Timestamp: 1446170809
\end{verbatim}

\end{quote}
Note that Jython already features an incomplete ctypes-module based on JFFI (which is part of \cite{JNR}). Without JyNI the output would look as follows:%
\begin{quote}\begin{verbatim}
<ctypes.CDLL instance at 0x2>
<ctypes._Function object at 0x3>
Traceback (most recent call last):
  File "/home/stefan/workspace/JyNI/JyNI-Demo/src
          /JyNIctypesTest.py", line 68, in <module>
  System.out.println(libc.time(0))
NotImplementedError: variadic functions not
supported yet;  specify a parameter list
\end{verbatim}

\end{quote}
JyNI bundles a custom version of \texttt{ctypes/\_\_init\_\_.py} and overrides the original one at import time. For the C-part JyNI can utilize the compiled \texttt{\_ctypes.so} file bundled with CPython (remember that JyNI is binary compatible to such libraries). In our example we make CPython's C extension folder available by appending its usual POSIX location \texttt{/usr/lib/python2.7/lib-dynload} to \texttt{sys.path}.

In \texttt{ctypes/\_\_init\_\_.py} we had to fix POSIX-recognition; it was based on \texttt{os.name}, which always reads “java” in Jython, breaking the original logic. See \cite{JyNI_GSoC} for details.

\subsection{Experimental NumPy-import%
  \label{experimental-numpy-import}%
}

As of JyNI alpha.4 it is possible to import NumPy 1.12 (current repository version, not yet released) and perform some very basic operations:\begin{Verbatim}[commandchars=\\\{\},fontsize=\footnotesize]
\PY{k+kn}{import} \PY{n+nn}{sys}
\PY{n}{sys}\PY{o}{.}\PY{n}{path}\PY{o}{.}\PY{n}{append}\PY{p}{(}\PY{l+s+s1}{\PYZsq{}}\PY{l+s+s1}{path\PYZus{}to\PYZus{}numpy1.12\PYZus{}pre\PYZhy{}release}\PY{l+s+s1}{\PYZsq{}}\PY{p}{)}
\PY{k+kn}{import} \PY{n+nn}{numpy} \PY{k+kn}{as} \PY{n+nn}{np}

\PY{n}{a} \PY{o}{=} \PY{n}{np}\PY{o}{.}\PY{n}{array}\PY{p}{(}\PY{p}{[}\PY{l+m+mi}{2}\PY{p}{,} \PY{l+m+mi}{5}\PY{p}{,} \PY{l+m+mi}{7}\PY{p}{]}\PY{p}{)}

\PY{k}{print} \PY{n}{a}
\PY{k}{print} \PY{l+m+mi}{3}\PY{o}{*}\PY{n}{a}
\PY{k}{print} \PY{n}{np}\PY{o}{.}\PY{n}{outer}\PY{p}{(}\PY{n}{a}\PY{p}{,} \PY{n}{a}\PY{p}{)}
\end{Verbatim}
This example yields the expected output:\begin{Verbatim}[commandchars=\\\{\},fontsize=\footnotesize]
\PY{p}{[}\PY{l+m+mi}{2} \PY{l+m+mi}{5} \PY{l+m+mi}{7}\PY{p}{]}
\PY{p}{[}\PY{l+m+mi}{6} \PY{l+m+mi}{15} \PY{l+m+mi}{21}\PY{p}{]}
\PY{p}{[}\PY{p}{[}\PY{l+m+mi}{4} \PY{l+m+mi}{10} \PY{l+m+mi}{14}\PY{p}{]}
 \PY{p}{[}\PY{l+m+mi}{10} \PY{l+m+mi}{25} \PY{l+m+mi}{35}\PY{p}{]}
 \PY{p}{[}\PY{l+m+mi}{14} \PY{l+m+mi}{35} \PY{l+m+mi}{49}\PY{p}{]}\PY{p}{]}
\end{Verbatim}
E.g. operations involving floating point numbers fail as of this writing, so a usable support is still in some distance. Considering its vast complexity, mastering NumPy's import-script \DUroletitlereference{numpy/\_\_init\_\_.py} is yet a crucial milestone on this front.

Major challenges here were dependencies on several other extensions. E.g. NumPy depends on ctypes and datetime (i.e. on datetime's C-API, rendering the Jython-variant of datetime insufficient) – see ctypes example in \DUrole{ref}{using-native-ctypes} and datetime example in \cite{JyNI_EP13}.

NumPy also contains Cython-generated code, which results in huge and hard to read C source-files. E.g. \DUroletitlereference{mtrand.c} from NumPy's random module contains more than 40.000 lines of code, making it a true challenge to debug. In this sense the feasible NumPy-import also demonstrates a basic Cython-capability of JyNI alpha.4.

\section{Roadmap%
  \label{roadmap}%
}

While NumPy- and SciPy-support is the driving motivation for the JyNI-project (since these extensions are of most scientific importance), it is very hard to assess how much work will be needed to actually reach this goal. Determining even the feasibility of full NumPy support beforehand would be a project for itself, so we focus on taking one  hurdle by another, converging to full NumPy support as far as possible.

Another important goal is better cross-platform support. Currently JyNI works (i.e. is tested) with Linux and OSX and hypothetically (i.e. untested) with other POSIX systems.

The planned main features for the next release (alpha.5) are support for the buffer protocol and improved NumPy-support. More testing with Cython and improved support will be addressed after JyNI alpha.5.

\end{document}